\documentclass[pre,aps,amsmath,amssymb,reprint,superscriptaddress]{revtex4-2}

\usepackage{graphicx}
\usepackage{dcolumn}
\usepackage{bm}
\usepackage[
    bookmarks=false,
    colorlinks=true,
    linkcolor=blue,
    citecolor=blue,
    urlcolor=blue
]{hyperref}
\usepackage{orcidlink}

\newcommand\ord{\mathop{\mathrm{ord}}}
\newcommand\erf{\mathop{\mathrm{erf}}}
\newcommand{\appropto}{\mathrel{\vcenter{
  \offinterlineskip\halign{\hfil$##$\cr
    \propto\cr\noalign{\kern2pt}\sim\cr\noalign{\kern-2pt}}}}}

\begin{document}

\title{Freely-jointed chain models with extensible links}
\author{Michael R. Buche \orcidlink{0000-0003-1892-0502}}
\email{mrbuche@sandia.gov}
\affiliation{Materials and Failure Modeling, Sandia National Laboratories, Albuquerque, NM 87185, USA}
\author{Meredith N. Silberstein \orcidlink{0000-0002-6853-9796}}
\affiliation{Sibley School of Mechanical and Aerospace Engineering, Cornell University, Ithaca, NY 14853, USA}
\author{Scott J. Grutzik \orcidlink{0000-0002-6490-3941}}
\affiliation{Materials and Failure Modeling, Sandia National Laboratories, Albuquerque, NM 87185, USA}
\date{\today}

\begin{abstract}
Analytical relations for the mechanical response of single polymer chains are valuable for modeling purposes, on both the molecular and continuum scale.
These relations can be obtained using statistical thermodynamics and an idealized single-chain model, such as the freely-jointed chain model.
In order to include bond stretching, the rigid links in the freely-jointed chain model can be made extensible, but this almost always renders the model analytically intractable.
Here, an asymptotically-correct statistical thermodynamic theory is used to develop analytic approximations for the single-chain mechanical response of this model.
The accuracy of these approximations is demonstrated using several link potential energy functions.
This approach can be applied to other single-chain models, and to molecular stretching in general.

\noindent DOI: \href{https://doi.org/10.1103/PhysRevE.106.024502}{10.1103/PhysRevE.106.024502}.
\end{abstract}

\maketitle

\section{Introduction}

The mechanical response of a single polymer chain can be obtained by measuring the end-to-end length as a function of an applied force.
For small forces, this single-chain mechanical response is primarily due to the reduction in entropy as the chain is extended \cite{treloar1949physics}.
Idealized single-chain models, such as the freely-jointed chain model, allow quantification of these physics.
The freely-joined chain (FJC) model consists of a series of rigid links connected by penalty-free hinges \cite{flory1969statistical}.
Using statistical thermodynamics \cite{mcq}, the single-chain mechanical response can be obtained exactly and closed-form in terms of the Langevin function \cite{rubinstein2003polymer}.
For large forces, bonds would begin to stretch in the real chain, so the rigid links of the FJC model should be made extensible using some potential energy function \cite{buche2021chain}.
Though the same thermodynamic principles apply \cite{buche2020statistical}, the necessary configuration integrals almost always become analytically intractable.
Currently, the only known exactly solvable model is the particular case of harmonic link potentials \cite{balabaev2009extension,manca2012elasticity}.
This is unfortunate, since exact relations enable more efficient modeling and a deeper fundamental understanding.
Analytic approximations are a good alternative, since they are efficient and are often quite accurate.

There are a variety of approaches that have been developed to approximate the single-chain mechanical response of freely-jointed chains with extensible links.
The simplest approach is to directly modify the Langevin function of the FJC single-chain mechanical response in order to yield the correct high-force behavior for a given link stiffness \cite{smith1996over}.
This approach has become especially popular in capturing data from single-chain pulling experiments, such as those involving DNA, with success largely enabled by high link stiffness \cite{oesterhelt1999,grebikova2016350,grebikova2014single,smith1996over,rief1997single,janshoff2000force,frey2012understanding,WANG19971335,wang2001single,calderon2008quantifying,McCauley2007,Bosco2013,camunas2016}.
Additional terms can be included to obtain an improved approximation for harmonic potentials \cite{balabaev2009extension,fiasconaro2019analytical}, enabling better accuracy at lower link stiffnesses and therefore more robust modeling \cite{radiom2017influence,buche2020statistical}.
This simplest approach can be generalized for anharmonic link potentials in order to capture the mechanical response up until the chain breaks, which is useful for large-deformation polymer network constitutive models \cite{buche2021chain,jayathilaka2021force}.
An alternative approach has been developed by \citet{mao2017rupture}, where a constructed free energy function is minimized with respect to link length in order to obtain an effective link length, and subsequently, the single-chain mechanical response.
This approach has been utilized quite frequently in polymer network constitutive models, using both harmonic \cite{talamini2018progressive, mao2018theory, li2020variational, mulderrig2021affine, arunachala2021energy, lamont2021rate} and anharmonic \cite{mao2017rupture, arora2020fracture, yang2020multiscale, xiao2021modeling, lavoie2019modeling, guo2021micromechanics, zhao2021multiscale, arora2021coarse} link potential energy functions, though it is heuristic to minimize thermodynamic free energies with respect to phase space degrees of freedom \cite{buche2021chain}.

Despite this effort and progress, a more complete approach of approximating the single-chain mechanical response of freely-jointed chains with extensible links is still needed.
Critically, there are currently no approaches that are demonstrably accurate in a well-understood regime of model parameters.
Further, any reliable approach should begin from and closely adhere to the principles of statistical thermodynamics.
Here, such an approach is developed using an asymptotically-correct statistical thermodynamic theory \cite{buche2021fundamental}.
Beginning from the partition function, relations for the single-chain mechanical response are obtained which are asymptotically valid as the link potentials become steep.
A potential is considered to be steep when both the potential energy scale and stiffness is large compared to thermal energy.
These relations are compared to existing relations in the literature, and additional useful relations are provided in the Appendix, such as that for the Helmholtz free energy.
The accuracy of the asymptotic approximations is then demonstrated using popular potential energy functions for the links -- harmonic, Morse \cite{morse1929diatomic}, log-squared \cite{mao2017rupture}, and Lennard-Jones \cite{jones1924determinationii} potentials -- where it is shown in each case that the approximations become accurate as the link potential becomes steep.
The model has been implemented in the open-source \texttt{Python} package \texttt{ufjc} \cite{buchegrutzikufjc2022}, which offers additional functionalities not shown here.

\section{Theory}

The freely-jointed chain (FJC) model consists of $N_b$ rigid links of length $\ell_b$; the links may pass through each other or overlap, and are connected in series by penalty-free hinges \cite{treloar1949physics,flory1969statistical,rubinstein2003polymer}.
This single-chain model is generalized to the $u$FJC model through assigning some potential energy function $u$ to each link and allowing the link length $\ell$ to fluctuate away from its rest length $\ell_b$ \cite{buche2021chain}.
Here the isotensional ensemble is considered, where a fixed force $f$ is applied to the chain and the expected chain end-to-end length $\xi$ is calculated using the partition function \cite{manca2012elasticity,fiasconaro2019analytical,buche2020statistical}.
The temperature $T$ is also fixed, or equivalently $\beta=1/k_\mathrm{B}T$ is fixed, where $k_\mathrm{B}$ is the Boltzmann constant.
Asymptotically-correct relations, valid for steep link potentials \cite{buche2021fundamental}, are obtained for the isotensional partition function and are subsequently used to obtain the isotensional mechanical response.
Steep potentials are characterized by large scale and stiffness compared with thermal energy, i.e. steep potentials are both deep and narrow.
An asymptotic relation for low to intermediate forces is first obtained, then another for high forces, and finally the two are matched in a composite relation for all forces.
A reduced form of this full relation is provided, which also becomes accurate in the limit of sufficiently steep link potentials.

\subsection{Low-to-intermediate force asymptotics}

First, consider the cases when the nondimensional force $\eta=\beta f\ell_b$ is small compared to the nondimensional link potential energy scale $\varepsilon=\beta u_c$, as well as the nondimensional link stiffness $\kappa=\beta\ell_b^2 u''(\ell_b)$.
Note that $u_c$ is the characteristic energy scale for the potential $u$, and that apostrophes denote derivatives, i.e. $u''=d^2u/d\ell^2$.
This is succinctly stated as $\eta\ll\varepsilon,\kappa$ and encompass both the low ($\eta<1$) and intermediate ($1<\eta\ll\varepsilon,\kappa$) force regimes, where the link potential is assumed to be steep ($\varepsilon,\kappa\gg 1$).
An asymptotic relation for the single-chain mechanical response $\gamma(\eta)$ is desired for this force regime, where $\gamma=\xi/N_b\ell_b$ is the nondimensional end-to-end length.
Only the single-link isotensional partition function ($\theta$ is the angle between the link and the force),

\begin{equation}
	\mathfrak{z}(f) = 
	\int e^{\beta f\ell\cos\theta} e^{-\beta u(\ell)} \,d^3\boldsymbol{\ell}
	,
\end{equation}
is necessary to obtain the single-chain mechanical response of the $u$FJC model since the link degrees of freedom become decoupled in the isotensional ensemble \cite{fiasconaro2019analytical,buche2020statistical,buche2021chain}.
After computing the angular integrals and nondimensionalizing the integrand, the result is

\begin{equation}\label{z1dintegral}
	\mathfrak{z}(\eta) = 
	4\pi\ell_b^3 \int \frac{\sinh(s\eta)}{s\eta} \, e^{-\varepsilon\phi(s)} \,s^2\,ds
	,
\end{equation}
where $\phi(s)\equiv\beta u(\ell_b s)/\varepsilon$ is the scaled nondimensional potential energy function, and $s$ is a dummy variable of integration.
This full-system partition function can be rewritten as an integral transform of the reference system partition function,

\begin{equation}\label{zb4asymp}
	\mathfrak{z}(\eta) = 
	\ell_b \int \mathfrak{z}_0(\eta, s) \, e^{-\varepsilon\phi(s)} \,ds
	,
\end{equation}
where the reference system partition function $\mathfrak{z}_0(\eta, s)$ is that of the FJC model with a link length of $s\ell_b$,

\begin{equation}\label{z0}
	\mathfrak{z}_0(\eta, s) = 
	4\pi\ell_b^2 \, \frac{\sinh(s\eta)}{s\eta}\,s^2
	.
\end{equation}
The expected FJC partition function \cite{rubinstein2003polymer} is obtained for $s=1$, and correspondingly, $\mathfrak{z}_0(\eta, 1)\equiv \mathfrak{z}_0(\eta)$ is defined.
Following \citet{buche2021fundamental}, an asymptotic approximation for Eq.~\eqref{zb4asymp} is now obtained in order to represent the full system in terms of the reference system and small corrections.
Since absolute free energies will not be required here, $\phi(1)=0$ is assumed without loss of generality in order to simplify results.
Assuming that $\phi(s)$ achieves a unique, hyperbolic minimum at the link rest length $s=1$, i.e. $\phi'(1)=0$ and $\phi''(1)>0$, it has the Taylor series expansion near $s=1$ given by

\begin{align}\label{taylor-phi}
	\phi(s) =& \
	\frac{1}{2}\,\phi''(1)(s-1)^2
	+ \frac{1}{6}\,\phi'''(1)(s-1)^3
	\nonumber \\ &
	+ \frac{1}{24}\,\phi''''(1)(s-1)^4 + \cdots
	.
\end{align}
The Taylor series of $\mathfrak{z}_0(\eta, s)$ about the same point is

\begin{align}\label{taylor-z0}
	\mathfrak{z}_0(\eta, s) =& \
	\mathfrak{z}_0(\eta,1)
	+ \mathfrak{z}_0'(\eta,1)(s - 1)
	\nonumber \\ &
	+ \frac{1}{2}\,\mathfrak{z}_0''(\eta,1)(s - 1)^2 + \cdots
	,
\end{align}
where $\mathfrak{z}_0(\eta,1)\equiv\mathfrak{z}_0(\eta)$ is given by Eq.~\eqref{z0}, and allows the $a$th derivative to be computed using the relation

\begin{equation}\label{taylor-exp}
	\left.\frac{\partial^a\mathfrak{z}_0(\eta, s)}{\partial s^a}\right|_{s=1} =
	\big[\eta^a + a\eta\coth(\eta)\big] \mathfrak{z}_0(\eta)
	.
\end{equation}
Laplace's method for approximating integrals \cite{bleistein1975asymptotic} is now applied to Eq.~\eqref{zb4asymp}.
For $\varepsilon\gg 1$, $e^{-\varepsilon\phi(s)}$ decays extremely rapidly away from $s=1$, such that Eq.~\eqref{zb4asymp} is reasonably approximated when expanding the integrand about $s=1$.
Accordingly, within Eq.~\eqref{zb4asymp}, $\mathfrak{z}_0(\eta,s)$ is given by Eq.~\eqref{taylor-z0}, $\phi(s)$ is given by Eq.~\eqref{taylor-phi}, and $e^{-\varepsilon\phi(s)}$ is given by the Gaussian function \cite{bender2013advanced}

\begin{align}
	e^{-\varepsilon\phi(s)} \sim
	e^{-\kappa(s-1)^2/2}\left\{
		1 + \varepsilon\left[
			\frac{\phi'''(1)}{6}(s-1)^3
		\right.\right. \nonumber \\ \left.\left.
			+ \frac{\phi''''(1)}{24}(s-1)^4
		\right]
		+ \frac{\varepsilon^2}{72}\left[\phi'''(1)\right]^2(s-1)^6 + \cdots
	\right\}
	,
\end{align}
where $\kappa\equiv\varepsilon\phi''(1)$ is the nondimensional link stiffness.
Using Eqs.~\eqref{taylor-phi}--\eqref{taylor-exp}, and the substitution $t=\sqrt{\kappa}(s-1)$, for $\varepsilon\gg 1$ the partition function in Eq.~\eqref{zb4asymp} is given by the asymptotic relation

\begin{equation}\label{zasympstep1}
	\mathfrak{z}(\eta) \sim
	\frac{\ell_b}{\sqrt{\kappa}} \, \mathfrak{z}_0(\eta) \int \, e^{-t^2/2} \left[1 + \frac{f(\eta,t)}{\kappa}\right] \,dt
	,
\end{equation}
where the function $f(\eta,t)$ is defined as

\begin{align}\label{f-asymp}
	f(\eta,t) \equiv& \
	t^2\left[\frac{\eta^2}{2} + \eta\coth(\eta)\right]
	- \frac{t^4}{6}\frac{\phi'''(1)}{2\phi''(1)}\left[1 + \eta\coth(\eta)\right]
	\nonumber \\ &
	+ \frac{t^6}{72}\left[\frac{\phi'''(1)}{\phi''(1)}\right]^2
	- \frac{t^4}{24}\frac{\phi''''(1)}{\phi''(1)}
	+ \cdots
	.
\end{align}
The ellipsis here represents terms that are odd power of $t$, which will not contribute to the Gaussian integrals in Eq.~\eqref{zasympstep1}, as well as terms that are $\ord(\kappa^{-1})$ or higher \cite{bender2013advanced}.
These higher order terms are now neglected, tantamount to making an additional asymptotic approximation based on $\kappa\gg 1$.
Notably, assumptions based on both $\varepsilon\gg 1$ and $\kappa\gg 1$ have now been incorporated, consistent with the steep potential requirement emphasized in this work.
Substituting Eq.~\eqref{f-asymp} into Eq.~\eqref{zasympstep1} and computing the resulting Gaussian integrals then yields

\begin{equation}\label{zasymporig}
	\mathfrak{z}(\eta) \sim
	\ell_b \sqrt{\frac{2\pi}{\kappa}} \, \mathfrak{z}_0(\eta) \left[1 + \frac{h(\eta)}{\kappa}\right]
	,
\end{equation}
where the correction function $h(\eta)$ is given by

\begin{align}\label{correction-function}
	h(\eta) \equiv& \
	\frac{\eta^2}{2} + \eta\coth(\eta) - \frac{\phi'''(1)}{2\phi''(1)}\left[1 + \eta\coth(\eta)\right]
	\nonumber \\ &
	+ \frac{5}{24}\left[\frac{\phi'''(1)}{\phi''(1)}\right]^2 - \frac{\phi''''(1)}{8\phi''(1)}
	.
\end{align}
Eq.~\eqref{zasymporig} represents an aproximation of the statistical thermodynamics of the full system (the $u$FJC model) in terms of that of the reference system (the FJC model), and is asymptotically-valid for stiff potentials ($\varepsilon,\kappa\gg 1$).
Again, since absolute free energies will not be required here, the last two terms in Eq.~\eqref{correction-function} are neglected without loss of generality.
When taking the logarithm of Eq.~\eqref{zasymporig}, which is needed for the free energy and subsequently the mechanical response, it is expedient to use another asymptotic approximation \cite{buche2021fundamental}.
Applying $\kappa\gg 1$ again, these would take the general form $\ln(1+x/\kappa)\sim x/\kappa$, a truncation of the Mercator series.
This is then used to make the final asymptotic approximation that

\begin{equation}\label{magic?}
	\left[1 + \frac{h(\eta)}{\kappa} \right] \sim
	e^{\eta^2/2\kappa}\left[1 + \frac{\eta}{c\kappa}\,\coth(\eta)\right]
	,
\end{equation}
where $1/c\equiv 1 - \phi'''(1)/2\phi''(1)$.
As will become clear later in Sec.~\ref{its-a-match!}, this step is necessary to ensure that the low-to-intermediate and high-force regimes are readily matched, all the while remaining consistent with the assumptions that lead to the preceeding asymptotic relation.
Substituting Eq.~\eqref{magic?} into Eq.~\eqref{zasymporig} obtains

\begin{equation}\label{zasymp}
	\mathfrak{z}(\eta) \sim
	\ell_b \sqrt{\frac{2\pi}{\kappa}} \, \mathfrak{z}_0(\eta) \, e^{\eta^2/2\kappa}\left[1 + \frac{\eta}{c\kappa}\,\coth(\eta)\right]
	,
\end{equation}
where $\mathfrak{z}_0(\eta)$ is given by Eq.~\eqref{z0}.
The isotensional single-chain mechanical response is given by $\gamma(\eta)=\partial\ln\mathfrak{z}(\eta)/\partial\eta$ \cite{manca2012elasticity,buche2020statistical}, so the corresponding asymptotic relation for $\gamma(\eta)$ is then obtained to be

\begin{equation}\label{gamma-low-to-intermediate-eta}
	\gamma(\eta) \sim
	\mathcal{L}(\eta) + \frac{\eta}{\kappa}\left[\frac{1 - \mathcal{L}(\eta)\coth(\eta)}{c + (\eta/\kappa)\coth(\eta)}\right] + \frac{\eta}{\kappa}
	.
\end{equation}
Eq.~\eqref{gamma-low-to-intermediate-eta} is asymptotically-valid for steep link potentials ($\varepsilon,\kappa\gg 1$) in the low-to-intermediate force regime (from $\eta<1$ to $1<\eta\ll\varepsilon,\kappa$), where $\mathcal{L}(\eta)=\coth(\eta)-1/\eta$ is the Langevin function.
Note that Eq.~\eqref{gamma-low-to-intermediate-eta}, in the case of harmonic potentials ($c=1$), has been obtained previously using other approaches \cite{balabaev2009extension,fiasconaro2019analytical}, but here it has been effectively generalized to account for anharmonicity.

\subsection{High-force asymptotics}

Next consider the high-force regime, where the nondimensional force is on the order of $\varepsilon$, the nondimensional link potential energy scale ($\eta=\ord(\varepsilon)$); the link potential is still assumed to be steep ($\varepsilon,\kappa\gg 1$).
These limits are applied in reconsidering the isotensional partition function, taking $\sinh(\eta)\sim e^\eta$ and the scaled nondimensional force $\tau\equiv\eta/\varepsilon$, obtaining

\begin{equation}\label{zeqnhgihf}
	\mathfrak{z}(\eta) \sim
	\frac{2\pi\ell_b^3}{\eta} \int e^{-\varepsilon\left[\phi(s) - \tau s\right]} \,s\,ds
	.
\end{equation}
Note that one is now effectively working with the scaled nondimensional total potential energy $\phi(s)-\tau s$, i.e. the potential energy minus the work, and that the reference system is now a one-dimensional array of links in series.
Also note that if only $\gamma(\eta)$ is desired, the asymptotic approach can be applied either before or after computing $\gamma(\eta)=\partial\ln\mathfrak{z}(\eta)/\partial\eta$, which for Eq.~\eqref{zeqnhgihf} is

\begin{equation}\label{ghighetaorig}
	\gamma(\eta) \sim
	\frac{\int e^{-\varepsilon\left[\phi(s) - \tau s\right]} \,s^2\,ds}
	{\int e^{-\varepsilon\left[\phi(s) - \tau s\right]} \,s\,ds} - \frac{1}{\eta}
	.
\end{equation}
The same asymptotic approach \cite{buche2021fundamental} used to get Eq.~\eqref{zasymporig} is now applied to each of the integrals in Eq.~\eqref{ghighetaorig}.
The total potential energy is expanded about the unique, hyperbolic minimum at the link stretch $s=\lambda$, found via

\begin{equation}\label{min-totPE}
	\frac{\partial}{\partial s}\left[\phi(s) - \tau s\right]_{s=\lambda} = 0
	,
\end{equation}
i.e. the solution of $\phi'(\lambda)=\tau=\eta/\varepsilon$.
Note that the unique minimum assumption prevents the asymptotic approach developed here from being immediately applicable to cases where Eq.~\eqref{min-totPE} has multiple solutions \cite{rief1998elastically,manca2013two,giordano2018helmholtz}.
However, once a method of choosing from or transitioning between particular solutions has been established \cite{giordano2017spin,benedito2018thermodynamics,benedito2018isotensional}, the asymptotic approach could be applied separately to each particular solution.
Now, applying the same asymptotic approach from the previous section to either $(a=1,2)$ of the integrals in Eq.~\eqref{ghighetaorig} about $s=\lambda$,

\begin{equation}\label{urgh}
	\int e^{-\varepsilon\left[\phi(s) - \tau s\right]} \,s^a\,ds \sim
	w(\eta) \lambda^a(\eta) \left[1 + \frac{\hat{h}_a(\eta)}{\hat{\kappa}(\eta)}\right]
	,
\end{equation}
where the instantaneous link stiffness is $\hat{\kappa}(\eta)\equiv\varepsilon\phi''[\lambda(\eta)]$.
The prefactor $w(\eta)$ in Eq.~\eqref{urgh} is

\begin{equation}
	w(\eta) \equiv
	\sqrt{\frac{2\pi}{\hat{\kappa}(\eta)}} \, e^{\eta\lambda(\eta)-\varepsilon\phi[\lambda(\eta)]}
	,
\end{equation}
and the correction functions $\hat{h}_a(\eta)$ in Eq.~\eqref{urgh} are

\begin{align}\label{blahblahblah}
	\hat{h}_a(\eta) \equiv& \
	\frac{a-1}{\lambda^2(\eta)} - \frac{\phi'''[\lambda(\eta)]}{2\phi''[\lambda(\eta)]} \frac{a}{\lambda(\eta)}
	\nonumber \\ &
	+ \frac{5}{24}\left\{\frac{\phi'''[\lambda(\eta)]}{\phi''[\lambda(\eta)]}\right\}^2 - \frac{\phi''''[\lambda(\eta)]}{8\phi''[\lambda(\eta)]}
	.
\end{align}
When computing the ratio in Eq.~\eqref{ghighetaorig} using the asymptotic relations in Eq.~\eqref{urgh}, the prefactor $w(\eta)$ cancels.
This is a benefit of computing $\gamma(\eta)=\partial\ln\mathfrak{z}(\eta)/\partial\eta$ before applying the asymptotic approach, and the result is

\begin{equation}\label{urghurgh}
	\gamma(\eta) \sim
	\lambda(\eta) + \frac{\lambda(\eta)\hat{h}_2(\eta)}{\hat{\kappa}(\eta) + \hat{h}_1(\eta)} - \frac{1}{\eta}
	.
\end{equation}
In the case of harmonic potentials ($h_2=\lambda^{-2}$, $h_1=0$, $\lambda=1+\eta/\kappa$) the second term in Eq.~\eqref{urghurgh} becomes $1/(\kappa+\eta)$, consistent with high-force approximations obtained previously \cite{fiasconaro2019analytical}.
In general, the second term in Eq.~\eqref{urghurgh} is quite complicated due to many terms of the form $\phi^{(a)}[\lambda(\eta)]$ via Eq.~\eqref{blahblahblah} and makes $\gamma(\eta)$ impractical.
In order to maintain the practicality of the asymptotic relations obtained here and facilitate matching in the next section, and since the second term in Eq.~\eqref{urghurgh} is a small correction, it is neglected, yielding

\begin{equation}\label{gamma-high-eta}
	\gamma(\eta) \sim
	\lambda(\eta) - \frac{1}{\eta}
	.
\end{equation}
Eq.~\eqref{gamma-high-eta} is asymptotically-valid for steep link potentials ($\varepsilon,\kappa\gg 1$) in the high-force regime ($\eta=\ord(\varepsilon)$).
Since $1/\eta$ is then small here, Eq.~\eqref{gamma-high-eta} could be simplified further to $\gamma(\eta) \sim \lambda(\eta)$, but this is not necessary.

\subsection{Matched asymptotics for all forces}\label{its-a-match!}

A composite asymptotic approximation for the $u$FJC isotensional single-chain mechanical response $\gamma(\eta)$ is now obtained, valid for all forces $\eta$ when the link potential is steep.
This asymptotic matching of Eqs.~\eqref{gamma-low-to-intermediate-eta} and \eqref{gamma-high-eta} is done using Prandtl's method \citep{powers2015mathematical}.
It is first verified that Eq.~\eqref{gamma-low-to-intermediate-eta} under $\eta\gg 1$ and Eq.~\eqref{gamma-high-eta} under $\eta\ll 1$ are equivalent, i.e. equal to $1-1/\eta+\eta/\kappa$.
To match, Eqs.~\eqref{gamma-low-to-intermediate-eta} and \eqref{gamma-high-eta} are added together and the common part $1-1/\eta+\eta/\kappa$ is subtracted.
Taking $\Delta\lambda\equiv\lambda-1$, the resulting composite asymptotic approximation is then

\begin{equation}\label{asymptotic}
	\gamma(\eta) \sim
	\mathcal{L}(\eta) + \frac{\eta}{\kappa}\left[\frac{1 - \mathcal{L}(\eta)\coth(\eta)}{c + (\eta/\kappa)\coth(\eta)}\right] + \Delta\lambda(\eta)
	,
\end{equation}
valid for steep link potentials ($\varepsilon,\kappa\gg 1$).
The asymptotic approximation in Eq.~\eqref{asymptotic} is the sum of three distinct terms:
the first term is the fully-entropic result for the FJC model \cite{treloar1949physics,flory1969statistical,rubinstein2003polymer};
the second term represents corrections related to rotation-vibration (entropic-enthalpic) coupling \cite{buche2021fundamental};
the third term is the fully-enthalpic incremental link stretch under a direct force $\eta$.
When Eq.~\eqref{asymptotic} is simplified in the case of a harmonic link potential ($c=1$, $\Delta\lambda=\eta/\kappa$), it matches an existing relation for the EFJC model \cite{balabaev2009extension,radiom2017influence,fiasconaro2019analytical}, which is highly accurate even for only moderately large $\kappa$ (as shown in Appendix~\ref{appefjcexact}, the error tends to be transcendentally small).
Here in Eq.~\eqref{asymptotic}, a more general asymptotic relation had been systematically obtained that handles the arbitrary link potentials of the $u$FJC model.

The reduced asymptotic approximation for the $u$FJC isotensional single-chain mechanical response,

\begin{equation}\label{simplest}
	\gamma(\eta) \sim \mathcal{L}(\eta) + \Delta\lambda(\eta)
	,
\end{equation}
is reached when $\kappa$ is sufficiently large and causes the second term in Eq.~\eqref{asymptotic} to contribute negligibly over all $\eta$.
Physically, this is equivalent to neglecting the coupling between link stretching and link rotation.
Eq.~\eqref{simplest} has been obtained previously using a combination of physical and mathematical arguments \cite{buche2021chain}, but here it has been more rigorously obtained.
When Eq.~\eqref{simplest} is simplified in the case of a harmonic link potential ($\Delta\lambda=\eta/\kappa$), it matches past expressions obtained heuristically \cite{oesterhelt1999,grebikova2016350,grebikova2014single}.
Since Eq.~\eqref{simplest} is asymptotic to $(1+\Delta\lambda)\mathcal{L}(\eta)$ for $\kappa\gg 1$, it also asymptotically matches many other expressions used for the harmonic case \cite{smith1996over,rief1997single,janshoff2000force,frey2012understanding,WANG19971335,wang2001single,calderon2008quantifying,McCauley2007,Bosco2013,camunas2016}.
The developments here verify these reduced relations for sufficiently steep potentials and offer applicability to arbitrary link potentials.
In Appendix~\ref{apphelm}, the reduced asymptotic relations for single-chain functions in the isometric ensemble (such as the Helmholtz free energy) are additionally provided using the Legendre transformation, which are asymptotically valid for a sufficiently large number of links \cite{manca2012elasticity,manca2014equivalence,buche2020statistical}.
This additional approximation is often necessary, since the FJC model is analytically complicated to solve in the isometric ensemble.
To avoid making the Legendre transformation approximation, one could use the FJC radial distribution function \cite{treloar1946statistical,wang1952statistical} to obtain the reference system partition function within a constant \cite{buche2020statistical} and apply the asymptotic approach, but the result would be highly impractical.

\section{Results}

The full asymptotic approach (Eq.~\eqref{asymptotic}) and the reduced asymptotic approach (Eq.~\eqref{simplest}) are now demonstrated in approximating the single-chain mechanical response of the $u$FJC model.
The harmonic link potential (the EFJC model) is considered first, followed by the log-squared potential \cite{mao2017rupture}, the Morse potential \cite{morse1929diatomic}, and the Lennard-Jones potential \cite{jones1924determinationii}.
For each, the link stretch is calculated using Eq.~\eqref{min-totPE}.
The asymptotic approaches are compared with an exact solution when available, and numerical quadrature otherwise.
Calculations were completed using the \texttt{Python} package \texttt{ufjc} \cite{buchegrutzikufjc2022}.

\subsection{Harmonic link potential}

\begin{figure}[t]
	\begin{center}
		\includegraphics{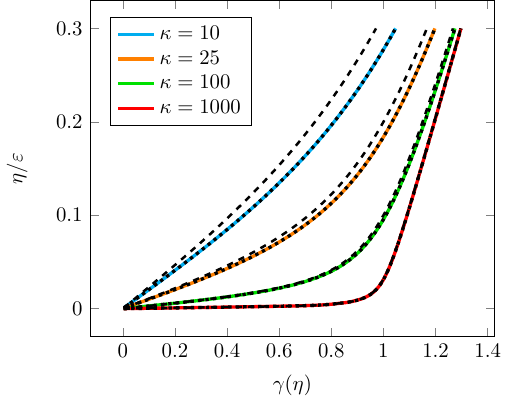}
	\end{center}
	\caption{\label{fig-efjc}
		The nondimensional single-chain mechanical response $\gamma(\eta)$ for the EFJC model, using the full asymptotic (dotted), reduced asymptotic (dashed), and exact (solid) approaches, for varying nondimensional link stiffness $\kappa=\varepsilon$.
	}
\end{figure}

\begin{figure}[t]
	\begin{center}
		\includegraphics{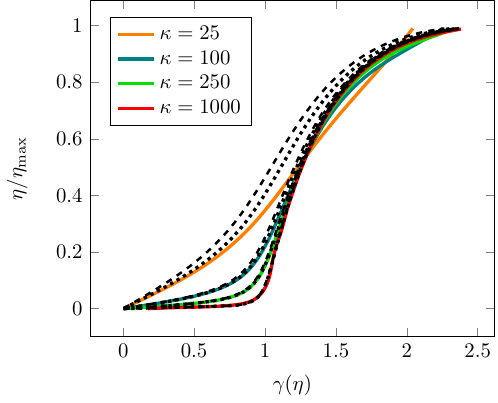}
	\end{center}
	\caption{\label{fig-log}
		The nondimensional single-chain mechanical response $\gamma(\eta)$ for the log-squared-FJC model, using the full asymptotic (dotted), reduced asymptotic (dashed), and quadrature (solid) approaches, for varying $\kappa=\varepsilon$.
	}
\end{figure}

Using harmonic link potentials with the $u$FJC model produces the EFJC model;
the scaled nondimensional potential energy function in this case is

\begin{equation}\label{phi-harmonic}
	\phi(s) = \frac{1}{2}(s-1)^2
	.
\end{equation}
Though the particularities may differ, the harmonic potential is the most common way of rendering the rigid links of the FJC model extensible \cite{manca2012elasticity, balabaev2009extension, radiom2017influence, talamini2018progressive, mao2018theory, fiasconaro2019analytical, buche2020statistical, li2020variational, jayathilaka2021force, mulderrig2021affine, arunachala2021energy, lamont2021rate, oesterhelt1999,grebikova2016350,grebikova2014single,smith1996over,rief1997single,janshoff2000force,frey2012understanding,WANG19971335,wang2001single,calderon2008quantifying,McCauley2007,Bosco2013,camunas2016}.
The full asymptotic, reduced asymptotic, and exact (see Appendix~\ref{appefjcexact}) approaches of obtaining the EFJC single-chain mechanical response $\gamma(\eta)$ are plotted in Fig.~\ref{fig-efjc} while varying the link stiffness $\kappa$.
For optimal readability, these results are given in terms of the scaled nondimensional force $\tau\equiv\eta/\varepsilon$.
The full asymptotic approach is negligibly different from the exact approach for all values of $\kappa$ considered; as shown in Appendix~\ref{appefjcexact}, this is due to the full asymptotic approximation being exactly correct to within transcendentally small terms in the case of harmonic links.
The reduced asymptotic approach tends to be inaccurate for moderate $\kappa$, but quickly becomes accurate for large $\kappa$.
Above $\kappa=100$, the difference between all three apparently vanishes, where the reduced asymptotic approach could be used in place of the exact approach for expediency since $\kappa$ is often larger than 100 when modeling experiments \cite{grebikova2014single, grebikova2016350, buche2021chain, jayathilaka2021force}.

\subsection{Other link potentials}

\begin{figure}[t]
	\begin{center}
		\includegraphics{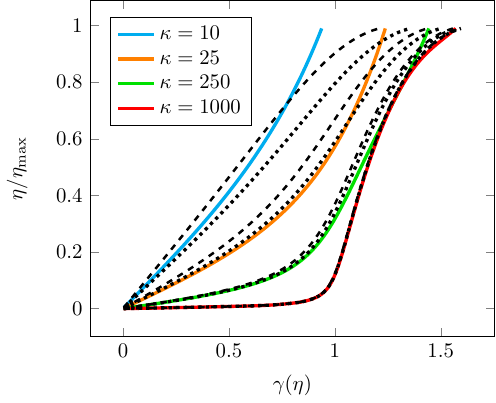}
	\end{center}
	\caption{\label{fig-morse}
		The nondimensional single-chain mechanical response $\gamma(\eta)$ for the Morse-FJC model ($\alpha=1$), using the full asymptotic (dotted), reduced asymptotic (dashed), and quadrature (solid) approaches, for varying $\kappa=2\alpha^2\varepsilon$.
	}
\end{figure}

When link stretches are expected to be large, the chosen link potential energy functions are generally anharmonic and escapable.
Similarly to using true strain in place of engineering strain in a continuum model, the harmonic potential can be replaced with the log-squared potential \cite{mao2017rupture, arora2020fracture, yang2020multiscale, xiao2021modeling, arora2021coarse}.
The scaled nondimensional potential energy function in this case is

\begin{equation}
	\phi(s) = \frac{1}{2}\big[\ln(s)\big]^2
	.
\end{equation}
Since the exact, analytic relation for $\gamma(\eta)$ is not known in this case (and the following cases), Eq.~\eqref{z1dintegral} is integrated using numerical quadrature \cite{numpy, scipy} in place of an exact approach.
The asymptotic approaches are compared with the quadrature results for varying $\kappa$ in Fig.~\ref{fig-log}, where $\eta$ is scaled by $\eta_\mathrm{max}=\varepsilon/e$.
While the full asymptotic approach does perform better (in matching the quadrature approach), neither asymptotic approach is especially accurate until $\kappa$ reaches 100 and above. 

The Morse potential \cite{morse1929diatomic} is another common choice for the link potential energy function of the $u$FJC model \cite{lavoie2019modeling, buche2021chain, guo2021micromechanics}.
The scaled nondimensional Morse potential energy function is

\begin{equation}
	\phi(s) = \left[1 - e^{-\alpha(s - 1)}\right]^2
	,
\end{equation}
where $\alpha$ is the Morse parameter, related to the nondimensional stiffness $\kappa=2\alpha^2\varepsilon$.
The asymptotic approaches are compared with the quadrature results for varying $\kappa$ in Fig.~\ref{fig-morse}, where $\eta$ is scaled by $\eta_\mathrm{max}=\alpha\varepsilon/2$.
Fig.~\ref{fig-morse} illustrates an important pathology of the quadrature approach that appears when dealing with escapable potentials, such as the Morse potential.
In order for the partition function in Eq.~\eqref{z1dintegral} to converge in quadrature methods, the integration must be constrained to prevent links from breaking.
For insufficiently steep link potentials, this constraint is non-physical and results in artificial strain-stiffening in the anharmonic regime, rather than the expected strain-softening of an escapable potential.
As the link potential becomes sufficiently steep, this pathology vanishes, and the expected behavior is obtained; all of this is seen clearly in Fig.~\ref{fig-morse}.
In short, any approach for breakable links is only valid when the link potentials are sufficiently steep.
Fig.~\ref{fig-morse} shows that the full asymptotic approach matches the quadrature approach more closely than the reduced asymptotic method does, and that all three methods converge as $\kappa$ becomes large.
In addition to being more interpretable and computationally expedient, note that the asymptotic methods do not suffer from the artificial strain-stiffening pathology.
It would then be best, in practice, to utilize the asymptotic approaches developed here when stretching breakable molecules.

\begin{figure}[t]
	\begin{center}
		\includegraphics{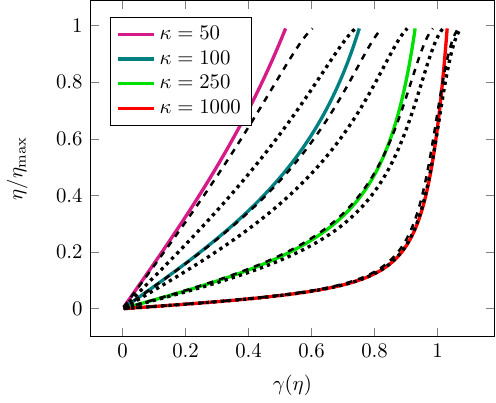}
	\end{center}
	\caption{\label{fig-Lennard-jones}
		The nondimensional single-chain mechanical response $\gamma(\eta)$ for the Lennard-Jones-FJC model, using the full asymptotic (dotted), reduced asymptotic (dashed), and quadrature (solid) approaches, for varying $\kappa=72\varepsilon$.
	}
\end{figure}

Lastly, the Lennard-Jones potential \cite{jones1924determinationii} can also be used as the link potential energy function \cite{zhao2021multiscale}.
The scaled nondimensional potential energy function in this case is

\begin{equation}
	\phi(s) = \frac{1}{s^{12}} - \frac{2}{s^6}
	.
\end{equation}
The asymptotic approaches are compared with the quadrature results for varying $\kappa$ in Fig.~\ref{fig-Lennard-jones}, where $\eta$ is scaled by $\eta_\mathrm{max}=12\varepsilon[(7/13)^{7/6} - (7/13)^{13/6}]$.
Each asymptotic approach converges for large $\kappa$, but interestingly, observe that the reduced asymptotic approach tends to match the quadrature approach more closely than the full asymptotic approach.
The Lennard-Jones potential is escapable, so the strain-stiffening pathology of the quadrature approach is also observed in Fig.~\ref{fig-Lennard-jones}; these two observations are related.
Since the nondimensional stiffness is nearly two orders of magnitude different from the nondimensional energy scale (i.e. $\kappa=72\varepsilon$), $\kappa$ must be quite large for $\varepsilon$ to also be sufficiently large to consider the potential steep.
This results in higher values of $\kappa$ being required for the approaches to converge in Fig.~\ref{fig-Lennard-jones} compared to the previous cases.
Note that the quadrature approach is not necessarily accurate for lower values of $\kappa$, so the reduced asymptotic approach matching more closely in Fig.~\ref{fig-Lennard-jones} could be misleading.
As shown next in Sec.~\ref{sec-error}, i.e. Fig.~\ref{fig-error}, the full asymptotic approach is actually more accurate in this case once $\kappa$ is sufficiently large.
To further analyze the steep potential requirement, one can consider an opposing case where the energy scale is high but the stiffness is low, or even zero: the square-well potential \cite{mcq}.
The asymptotic approach cannot be applied at all in this case, even though the results do approach that of the reference system (FJC) as the potential narrows (see Appendix~\ref{appsw}).

\subsection{Error analysis}\label{sec-error}

\begin{figure}[t]
	\begin{center}
		\includegraphics{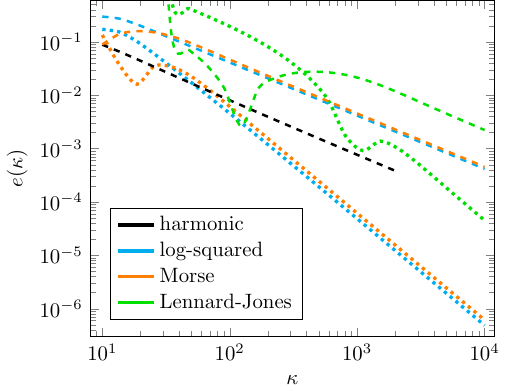}
	\end{center}
	\caption{\label{fig-error}
		The relative error $e$ as a function of the nondimensional stiffness $\kappa$, using the full asymptotic (dotted) and reduced asymptotic (dashed) approaches.
	}
\end{figure}

The accuracy of the full (Eq.~\eqref{asymptotic}) and reduced (Eq.~\eqref{simplest}) asymptotic approaches are now analyzed by computing the error relative to some baseline approach.
The $L_2$ norm is chosen in defining the relative error $e$, which will be computed while varying the nondimensional link stiffness $\kappa$.
This is given by

\begin{equation}
	e(\kappa) \equiv
	\sqrt{\dfrac{\int_0^{\eta_\mathrm{max}} \left[ \gamma(\eta) - \gamma_0(\eta) \right]^2 d\eta}{\int_0^{\eta_\mathrm{max}} \gamma_0^2(\eta) \,d\eta}}
	,
\end{equation}
where $\gamma_0(\eta)$ is the exact solution in the case of the harmonic link potential (see Appendix~\ref{appefjcexact}), and is given by the numerical quadrature otherwise.
The results for either asymptotic approach applied to the harmonic, log-squared, Morse, and Lennard-Jones potentials are shown in Fig.~\ref{fig-error}.
Note that $\eta_\mathrm{max}$ is chosen as $0.3\varepsilon$ for the harmonic link potential, consistent with Fig.~\ref{fig-efjc}.
Also note that the full asymptotic approach for the harmonic link potential is not shown, as the error is many orders of magnitude smaller due to it being correct to within transcendentally small terms (see Appendix~\ref{appefjcexact}).
Fig.~\ref{fig-error} shows in general that error trends can be somewhat unpredictable at first, but become quite predictable (in terms of slope) as $\kappa$ becomes large.
The unpredictability at lower values of $\kappa$ can be attributed, at least partially, to the inaccuracy of the quadrature approach in the same regime.
Other approaches for $\gamma_0(\eta)$ will have the same issue since any approach for breakable links is generally invalid in this regime.
This is most apparent when comparing the harmonic and Lennard-Jones potentials: the quadrature approach is exact for the inescapable harmonic potential, so $e(\kappa)$ in log-log is a predictable line.
Conversely, the quadrature approach only becomes accurate for the Lennard-Jones potential for very large $\kappa$ needed to ensure $\varepsilon=\kappa/72$ is also sufficiently large, so $e(\kappa)$ only becomes predictable above $\kappa=1000$.
For sufficiently large $\kappa$, Fig.~\ref{fig-error} shows that the relative error in the full and reduced asymptotic approaches have log-log slopes of $-2$ and $-1$, respectively.
These slopes seem to confirm that the full (Eq.~\eqref{asymptotic}) and reduced (Eq.~\eqref{simplest}) asymptotic approaches are correct within terms that are $\ord(\kappa^{-2})$ and $\ord(\kappa^{-1})$, respectively, for sufficiently steep potentials.

\section{Conclusion}

An asymptotically-correct statistical thermodynamic theory has been applied to develop analytic approximations for the single-chain mechanical response of freely jointed chains with extensible links, i.e. the $u$FJC model.
The full asymptotic relation contains both entropic and enthalpic contributions as well as the coupling between them; when this coupling is neglected, the reduced asymptotic relation is obtained.
These asymptotic relations are valid as the link potential energy functions become steep, meaning both the potential energy scale as well as the stiffness become large compared with thermal energy.
For escapable potentials, this steepness is also understood as the potential well being both deep and narrow.
These asymptotic approaches were verified by comparing with the exact, analytic approach in the case of harmonic link potentials, using both parametric study and mathematical analysis.
Parametric studies were performed for the log-squared, Morse, and Lennard-Jones potentials, where exact results were unavailable and a quadrature approach was used.
In each case, the asymptotic approaches became increasingly accurate as the potentials became steep.
It was observed that the quadrature method suffers from an artificial strain-stiffening pathology for escapable potentials before the steep limit is met, which encourages use of the more robust asymptotic approaches.
The success of this asymptotic approach as demonstrated here, using the freely-jointed chain model as a reference system, indicates probable success for general molecular stretching models.
While even reference systems are often analytically intractable, this asymptotic approach can still be applied on an approximation for the reference system to obtain one for the full system.

\begin{acknowledgments}
This material is based in part upon work supported by the National Science Foundation, United States under Grant No.~CAREER-1653059.
This work was supported by the Laboratory Directed Research and Development (LDRD) program at Sandia National Laboratories under project~222398.
Sandia National Laboratories is a multi-mission laboratory managed and operated by National Technology and Engineering Solutions of Sandia, LLC., a wholly owned subsidiary of Honeywell International, Inc., for the U.S. Department of Energy's National Nuclear Security Administration under contract DE-NA0003525.
This paper describes objective technical results and analysis.
Any subjective views or opinions that might be expressed in the paper do not necessarily represent the views of the U.S. Department of Energy or the United States Government.
The authors thank Roger F. Loring for helpful thoughts and discussion regarding this work.
\end{acknowledgments}

\appendix

\section{The harmonic link potential}\label{appefjcexact}

Here the exact isotensional partition function is obtained in the case of a harmonic link potential, i.e. in the case of the EFJC model.
This exact result is then rewritten in terms of the asymptotic relation for the isotensional partition function obtained in the manuscript.
The ratio of the exact and asymptotic result is shown to be equal to unity plus terms that tend to be transcendentally small, which explains the strong accuracy of the asymptotic theory for harmonic potentials.

The isotensional partition function in Eq.~\eqref{z1dintegral} can be integrated exactly for the harmonic link potential in Eq.~\eqref{phi-harmonic}.
Using a symbolic toolbox, such as \texttt{Mathematica} \cite{Mathematica}, this exact result can be obtained:

\begin{align}\label{z-efjc-exact}
	\mathfrak{z}(\eta) = & \
	\pi\ell_b^3\sqrt{\frac{2\pi}{\kappa}} \, \frac{e^{\eta^2/2\kappa}}{\eta}\left\{e^{\eta}\left(\frac{\eta}{\kappa} + 1\right)\left[1 + \erf\left(\frac{\eta+\kappa}{\sqrt{2\kappa}}\right)\right]
	\right. \nonumber \\ & ~~~~~~~~ \left.
	+ e^{-\eta}\left(\frac{\eta}{\kappa} - 1\right)\left[1 - \erf\left(\frac{\eta-\kappa}{\sqrt{2\kappa}}\right)\right]\right\}
	.
\end{align}
Eq.~\eqref{z-efjc-exact} can be utilized effectively \cite{manca2012elasticity,buchegrutzikufjc2022}, but it is often considered practically undesirable due to the presence of the error function \cite{balabaev2009extension,fiasconaro2019analytical}.
To simplify, Eq.~\eqref{z-efjc-exact} can be rewritten in terms of the asymptotic approximation in Eq.~\eqref{zasymp}.
After defining $g(\eta)$ as

\begin{equation}
	g(\eta) \equiv
	\frac{y^+(\eta) - y^-(\eta)}{4\sinh(\eta)\left[1 + (\eta/\kappa)\coth(\eta)\right]} - \frac{1}{2}
	,
\end{equation}
where either function $y^\pm(\eta)$ is defined as

\begin{equation}
	y^\pm(\eta) \equiv
	e^{\pm\eta}\left(\frac{\eta}{\kappa} \pm 1\right)\erf\left(\frac{\eta\pm\kappa}{\sqrt{2\kappa}}\right)
	,
\end{equation}
the exact result for $\mathfrak{z}(\eta)$ can be written in terms of $\mathfrak{z}_\mathrm{a}(\eta)$, given by Eq.~\eqref{zasymp} with $c=1$, as

\begin{equation}
	\mathfrak{z}(\eta) =
	\mathfrak{z}_\mathrm{a}(\eta) \left[1 + g(\eta)\right]
	,
\end{equation}
so that $g(\eta)$ is the correction to the asymptotic result.
Recall that $\varepsilon=\kappa$ for the EFJC model, so $\kappa\gg 1$ is considered alone.
For $\kappa\gg 1$ and small to intermediate forces $\eta\ll\kappa$, the error functions have the asymptotic relation due to their large arguments \cite{hinch1991perturbation}

\begin{align}
	\erf\left(\frac{\eta\pm\kappa}{\sqrt{2\kappa}}\right) \sim &
	\erf\left(\pm\sqrt{\frac{\kappa}{2}}\right) 
	\sim \pm 1 \mp \sqrt{\frac{2}{\pi\kappa}} \, e^{-\kappa/2}
	,
\end{align}
which, after simplifying, means that $g(\eta)$ is transcendentally small (exponentially small) for $\eta\ll\kappa$ and $\kappa\gg 1$:

\begin{equation}
	g(\eta) \sim
	-\frac{e^{-\kappa/2}}{\sqrt{2\pi\kappa}}
	\quad\text{for }\kappa\gg 1\text{ and }\eta\ll\kappa
	.
\end{equation}
For large forces $\eta=\ord(\kappa)$, $y^+(\eta)$ is transcendentally small due to the large argument of the error function (similar to the above), and $y^-(\eta)$ is transcendentally small due to $e^{-\eta}=e^{-\ord(\kappa)}$.
Further, $1/\sinh(\eta)\sim e^{-\ord(\kappa)}$ is also transcendentally small in this case.
Therefore $g(\eta)$ tends to be transcendentally small for all values of $\eta$ when $\kappa\gg 1$, showing why the asymptotic theory performs so well in the case of harmonic potentials, even when only moderately stiff.

\section{Approximate isometric ensemble}\label{apphelm}

Here the reduced asymptotic relation for the nondimensional Helmholtz free energy per link is obtained for the $u$FJC model, which is an isometric ensemble quantity.
This relation makes use of the reduced asymptotic relation for the isotensional single-chain mechanical response from the manuscript, as well as the Legendre transformation, and is then valid when $\kappa\gg 1$, $\varepsilon\gg 1$, and $N_b\gg 1$ are all simultaneously true.
This result is the used to obtain asymptotic relations for the equilibrium probability density distributions and reaction rate coefficient function in the isometric ensemble.

The Helmholtz free energy is given by $\psi(\gamma)=-(1/\beta)\ln\mathfrak{q}(\gamma)$, where $\mathfrak{q}(\gamma)$ is the partition function in the isometric ensemble \cite{buche2020statistical}.
The nondimensional Helmholtz free energy per link, $\vartheta(\gamma)=\beta\psi(\gamma)/N_b$, is desired.
The Legendre transformation method, asymptotically valid for sufficiently long chains \cite{manca2012elasticity,manca2014equivalence,buche2020statistical} and appreciable loads \cite{neumann2003precise,suzen2009ensemble}, allows one to write \cite{buche2021chain}

\begin{equation}
	\vartheta(\gamma) \sim \gamma\eta(\gamma) - \int \gamma(\eta) \,d\eta
	.
\end{equation}
Substitute in the reduced asymptotic approximation for $\gamma(\eta)$ in Eq.~\eqref{simplest}.
The entropic term (the Langevin function) produces the Helmholtz free energy for the FJC model \cite{buche2021chain}.
The enthalpic term (link stretching), after integrating by parts, produces the link potential energy.
The result is then

\begin{equation}\label{simplestvartheta}
	\vartheta(\gamma) \sim 
	\ln\left\{ \frac{\eta(\gamma)\exp\{\eta(\gamma)\mathcal{L}[\eta(\gamma)]\}}{\sinh[\eta(\gamma)]} \right\}
	+ \varepsilon\phi\{\lambda[\eta(\gamma)]\}
	.
\end{equation}
Note that this decoupling of the entropic and enthalpic contributions to the Helmholtz free energy is a product of combining both the asymptotic approach and the Legendre transformation method, and additionally, the decoupling itself is not a Legendre transformation.
The nondimensional potential energy $\varepsilon\phi=\beta u$ is a function of the link stretch $\lambda$, which is a function of the nondimensional force $\eta$ that would result from the chain being extended to a nondimensional end-to-end length of $\gamma$.
Recall that this $\eta$ is calculated as a function of $\gamma$ from inverting the isotensional $\gamma(\eta)$ in Eq.~\eqref{simplest}.
In what follows, $\eta$ will be written with the understanding that $\eta=\eta(\gamma)$, i.e. Eq.~\eqref{simplestvartheta} would be

\begin{equation}\label{simplestvartheta2}
	\vartheta(\gamma) \sim 
	\ln\left\{\frac{\eta\exp[\eta\mathcal{L}(\eta)]}{\sinh(\eta)}\right\}
	+ \varepsilon\phi[\lambda(\eta)]
	.
\end{equation}
This same relation was obtained by \citet{buche2021chain}, but now it has been arrived at more rigorously.
Eq.~\eqref{simplestvartheta2} has several notable features:
first, it has been obtained in methodical fashion beginning from the basic principles of statistical thermodynamics;
second, it is assuredly asymptotically valid in the limit of numerable and stiff links;
third, it separates the entropic contribution to the free energy from the potential energy, facilitating a polymer network constitutive model to allow the potential energy to govern chain rupture \cite{mao2017rupture}.

Using the asymptotic relation for $\vartheta(\gamma)$ in Eq.~\eqref{simplestvartheta2}, an asymptotic relation for $P_\mathrm{eq}(\gamma)\propto e^{-N_b\vartheta}$, the probability density distribution of end-to-end lengths at equilibrium, can be written \cite{buche2020statistical,buche2021chain}.
Since the probability density at equilibrium where stiff links are stretched will be quite small, potential energy terms can typically be neglected \cite{buche2020statistical,lamont2021rate}.
One then obtains the asymptotic relation

\begin{equation}
	P_\mathrm{eq}(\gamma) \appropto
	\left\{ \frac{\sinh(\eta)}{\eta\exp[\eta\mathcal{L}(\eta)]} \right\}^{N_b}
	,
\end{equation}
where $\eta=\eta(\gamma)$ is still evaluated using $\gamma(\eta)$ in Eq.~\eqref{simplest}.
The corresponding radial distribution function is $g_\mathrm{eq}(\gamma)=4\pi\gamma^2 P_\mathrm{eq}(\gamma)$.
If transition state theory is utilized to describe the rate of breaking one of the $N_b$ links in the chain, and entropic effects are neglected compared to dominating enthalpic effects \cite{buche2021chain}, an asymptotic relation for the reaction rate coefficient function is obtained,

\begin{equation}
	k(\gamma) \sim
	N_b k_0 e^{\varepsilon\phi[\lambda(\eta)]}
	,
\end{equation}
where $k_0$ is the initial rate of breaking for a single link.

\section{The square-well potential}\label{appsw}

\begin{figure}[t]
	\begin{center}
		\includegraphics{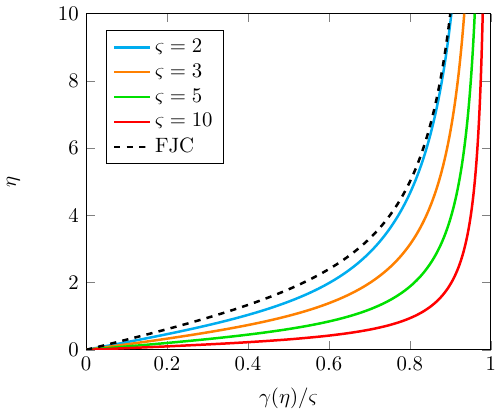}
	\end{center}
	\caption{\label{fig-sw}
	The nondimensional single-chain mechanical response $\gamma(\eta)$ scaled by $\varsigma$ for the SWFJC model.
	For $\varsigma\to 1$, the SWFJC model is equivalent to the FJC model.
	}
\end{figure}

Here the exact isotensional partition function and single-chain mechanical response is obtained in the case of a square-well link potential.
The scaled nondimensional potential energy function $\phi(s)=\beta u(\ell_b s)/\varepsilon$ for the square-well potential is given by

\begin{equation}\label{phi-sw}
	\phi(s) =
	\begin{cases}
	\infty, & s < 1, \\
	0, & 1\leq s < \varsigma, \\
	1, & s\geq\varsigma
	,
	\end{cases}
\end{equation}
where $(\varsigma-1)\ell_b$ is well width, $\varsigma\geq 1$, and $\varepsilon$ is the nondimensional well depth.
Consistent with the manuscript, the links are assumed not to break, which in this case bounds the integral in Eq.~\eqref{z1dintegral} to $s\in[1,\varsigma]$.
This can be computed exactly,

\begin{align}
	\mathfrak{z}(\eta) =&
	4\pi\ell_b^3 \int_1^\varsigma \frac{\sinh(s\eta)}{s\eta} \,s^2\,ds
	,\\ =&
	\frac{4\pi\ell_b^3}{\eta^3}\left[w(\eta,\varsigma) - w(\eta,1)\right]
	\label{z-swfjc}
	,
\end{align}
where $w(\eta,\varsigma)\equiv\varsigma\eta\cosh(\varsigma\eta)-\sinh(\varsigma\eta)$.
As the width of the well shrinks ($h\equiv \varsigma-1\to 0$), the square-well-FJC (SWFJC) model should become the FJC model.
This can be directly verified by scaling Eq.~\eqref{z-swfjc} by the scale of the lost degree of freedom $h\equiv \varsigma-1$ and taking the limit $h\to 0$, which yields Eq.~\eqref{z0} with $s=1$.
The isotensional single-chain mechanical response $\gamma(\eta)=\partial\ln\mathfrak{z}(\eta)/\partial\eta$ of the SW-FJC, using Eq.~\eqref{z-swfjc}, is exactly given by

\begin{equation}\label{gamma-swfjc}
	\gamma(\eta) =
	\frac{\varsigma^2\eta\sinh(\varsigma\eta) - \eta\sinh(\eta)}{w(\eta,\varsigma) - w(\eta,1)} - \frac{3}{\eta}
	.
\end{equation}
This result is plotted in Fig.~\ref{fig-sw} parameterized over $\varsigma$.
Applying the same limit $\varsigma\to 1$ here results in $\gamma(\eta)=\mathcal{L}(\eta)$, i.e. the FJC model, as expected.
Note that the single-chain mechanical response in Eq.~\eqref{gamma-swfjc} is independent of the well depth $\varepsilon$.
Additionally, note that there is no unique reference system since $\phi(s)$ in Eq.~\eqref{phi-sw} is minimized for a continuous distribution of arguments, which is because the square-well potential is flat and therefore has no stiffness.
For these reasons, the asymptotic approach developed in the manuscript cannot be applied to the square-well potential, or similar potentials.
This example helps illustrate the necessity of examining both the depth and the stiffness of a given link potential when determining whether the potential can be considered steep and the asymptotic approach can be applied.

\bibliography{main}

\begin{thebibliography}{60}%
\makeatletter
\providecommand \@ifxundefined [1]{%
 \@ifx{#1\undefined}
}%
\providecommand \@ifnum [1]{%
 \ifnum #1\expandafter \@firstoftwo
 \else \expandafter \@secondoftwo
 \fi
}%
\providecommand \@ifx [1]{%
 \ifx #1\expandafter \@firstoftwo
 \else \expandafter \@secondoftwo
 \fi
}%
\providecommand \natexlab [1]{#1}%
\providecommand \enquote  [1]{``#1''}%
\providecommand \bibnamefont  [1]{#1}%
\providecommand \bibfnamefont [1]{#1}%
\providecommand \citenamefont [1]{#1}%
\providecommand \href@noop [0]{\@secondoftwo}%
\providecommand \href [0]{\begingroup \@sanitize@url \@href}%
\providecommand \@href[1]{\@@startlink{#1}\@@href}%
\providecommand \@@href[1]{\endgroup#1\@@endlink}%
\providecommand \@sanitize@url [0]{\catcode `\\12\catcode `\$12\catcode
  `\&12\catcode `\#12\catcode `\^12\catcode `\_12\catcode `\%12\relax}%
\providecommand \@@startlink[1]{}%
\providecommand \@@endlink[0]{}%
\providecommand \url  [0]{\begingroup\@sanitize@url \@url }%
\providecommand \@url [1]{\endgroup\@href {#1}{\urlprefix }}%
\providecommand \urlprefix  [0]{URL }%
\providecommand \Eprint [0]{\href }%
\providecommand \doibase [0]{https://doi.org/}%
\providecommand \selectlanguage [0]{\@gobble}%
\providecommand \bibinfo  [0]{\@secondoftwo}%
\providecommand \bibfield  [0]{\@secondoftwo}%
\providecommand \translation [1]{[#1]}%
\providecommand \BibitemOpen [0]{}%
\providecommand \bibitemStop [0]{}%
\providecommand \bibitemNoStop [0]{.\EOS\space}%
\providecommand \EOS [0]{\spacefactor3000\relax}%
\providecommand \BibitemShut  [1]{\csname bibitem#1\endcsname}%
\let\auto@bib@innerbib\@empty
\bibitem [{\citenamefont {Treloar}(1949)}]{treloar1949physics}%
  \BibitemOpen
  \bibfield  {author} {\bibinfo {author} {\bibfnamefont {L.~R.~G.}\
  \bibnamefont {Treloar}},\ }\href
  {https://global.oup.com/academic/product/the-physics-of-rubber-elasticity-9780198570271}
  {\emph {\bibinfo {title} {The Physics of Rubber Elasticity}}}\ (\bibinfo
  {publisher} {Clarendon Press},\ \bibinfo {year} {1949})\BibitemShut {NoStop}%
\bibitem [{\citenamefont {Flory}(1969)}]{flory1969statistical}%
  \BibitemOpen
  \bibfield  {author} {\bibinfo {author} {\bibfnamefont {P.~J.}\ \bibnamefont
  {Flory}},\ }\href
  {https://scholar.google.com/scholar_lookup?title=Statistical%20Mechanics%20of%20Chain%20Molecules&author=P.J.%20Flory&publication_year=1969}
  {\emph {\bibinfo {title} {Statistical Mechanics of Chain Molecules}}}\
  (\bibinfo  {publisher} {Interscience},\ \bibinfo {year} {1969})\BibitemShut
  {NoStop}%
\bibitem [{\citenamefont {McQuarrie}(2000)}]{mcq}%
  \BibitemOpen
  \bibfield  {author} {\bibinfo {author} {\bibfnamefont {D.~A.}\ \bibnamefont
  {McQuarrie}},\ }\href
  {https://uscibooks.aip.org/books/statistical-mechanics/} {\emph {\bibinfo
  {title} {Statistical Mechanics}}}\ (\bibinfo  {publisher} {University Science
  Books},\ \bibinfo {year} {2000})\BibitemShut {NoStop}%
\bibitem [{\citenamefont {Rubinstein}\ and\ \citenamefont
  {Colby}(2003)}]{rubinstein2003polymer}%
  \BibitemOpen
  \bibfield  {author} {\bibinfo {author} {\bibfnamefont {M.}~\bibnamefont
  {Rubinstein}}\ and\ \bibinfo {author} {\bibfnamefont {R.~H.}\ \bibnamefont
  {Colby}},\ }\href
  {https://global.oup.com/academic/product/polymer-physics-9780198520597}
  {\emph {\bibinfo {title} {Polymer Physics}}}\ (\bibinfo  {publisher} {Oxford
  University Press},\ \bibinfo {year} {2003})\BibitemShut {NoStop}%
\bibitem [{\citenamefont {Buche}\ and\ \citenamefont
  {Silberstein}(2021)}]{buche2021chain}%
  \BibitemOpen
  \bibfield  {author} {\bibinfo {author} {\bibfnamefont {M.~R.}\ \bibnamefont
  {Buche}}\ and\ \bibinfo {author} {\bibfnamefont {M.~N.}\ \bibnamefont
  {Silberstein}},\ }\bibfield  {title} {\bibinfo {title} {Chain breaking in the
  statistical mechanical constitutive theory of polymer networks},\ }\href
  {https://doi.org/10.1016/j.jmps.2021.104593} {\bibfield  {journal} {\bibinfo
  {journal} {Journal of the Mechanics and Physics of Solids}\ }\textbf
  {\bibinfo {volume} {156}},\ \bibinfo {pages} {104593} (\bibinfo {year}
  {2021})}\BibitemShut {NoStop}%
\bibitem [{\citenamefont {Buche}\ and\ \citenamefont
  {Silberstein}(2020)}]{buche2020statistical}%
  \BibitemOpen
  \bibfield  {author} {\bibinfo {author} {\bibfnamefont {M.~R.}\ \bibnamefont
  {Buche}}\ and\ \bibinfo {author} {\bibfnamefont {M.~N.}\ \bibnamefont
  {Silberstein}},\ }\bibfield  {title} {\bibinfo {title} {Statistical
  mechanical constitutive theory of polymer networks: The inextricable links
  between distribution, behavior, and ensemble},\ }\href
  {https://doi.org/10.1103/PhysRevE.102.012501} {\bibfield  {journal} {\bibinfo
   {journal} {Physical Review E}\ }\textbf {\bibinfo {volume} {102}},\ \bibinfo
  {pages} {012501} (\bibinfo {year} {2020})}\BibitemShut {NoStop}%
\bibitem [{\citenamefont {Balabaev}\ and\ \citenamefont
  {Khazanovich}(2009)}]{balabaev2009extension}%
  \BibitemOpen
  \bibfield  {author} {\bibinfo {author} {\bibfnamefont {N.}~\bibnamefont
  {Balabaev}}\ and\ \bibinfo {author} {\bibfnamefont {T.}~\bibnamefont
  {Khazanovich}},\ }\bibfield  {title} {\bibinfo {title} {Extension of chains
  composed of freely joined elastic segments},\ }\href
  {https://doi.org/10.1134/S1990793109020109} {\bibfield  {journal} {\bibinfo
  {journal} {Russian Journal of Physical Chemistry B}\ }\textbf {\bibinfo
  {volume} {3}},\ \bibinfo {pages} {242} (\bibinfo {year} {2009})}\BibitemShut
  {NoStop}%
\bibitem [{\citenamefont {Manca}\ \emph {et~al.}(2012)\citenamefont {Manca},
  \citenamefont {Giordano}, \citenamefont {Palla}, \citenamefont {Zucca},
  \citenamefont {Cleri},\ and\ \citenamefont {Colombo}}]{manca2012elasticity}%
  \BibitemOpen
  \bibfield  {author} {\bibinfo {author} {\bibfnamefont {F.}~\bibnamefont
  {Manca}}, \bibinfo {author} {\bibfnamefont {S.}~\bibnamefont {Giordano}},
  \bibinfo {author} {\bibfnamefont {P.~L.}\ \bibnamefont {Palla}}, \bibinfo
  {author} {\bibfnamefont {R.}~\bibnamefont {Zucca}}, \bibinfo {author}
  {\bibfnamefont {F.}~\bibnamefont {Cleri}},\ and\ \bibinfo {author}
  {\bibfnamefont {L.}~\bibnamefont {Colombo}},\ }\bibfield  {title} {\bibinfo
  {title} {Elasticity of flexible and semiflexible polymers with extensible
  bonds in the {G}ibbs and {H}elmholtz ensembles},\ }\href
  {https://doi.org/10.1063/1.4704607} {\bibfield  {journal} {\bibinfo
  {journal} {The Journal of Chemical Physics}\ }\textbf {\bibinfo {volume}
  {136}},\ \bibinfo {pages} {154906} (\bibinfo {year} {2012})}\BibitemShut
  {NoStop}%
\bibitem [{\citenamefont {Smith}\ \emph {et~al.}(1996)\citenamefont {Smith},
  \citenamefont {Cui},\ and\ \citenamefont {Bustamante}}]{smith1996over}%
  \BibitemOpen
  \bibfield  {author} {\bibinfo {author} {\bibfnamefont {S.~B.}\ \bibnamefont
  {Smith}}, \bibinfo {author} {\bibfnamefont {Y.}~\bibnamefont {Cui}},\ and\
  \bibinfo {author} {\bibfnamefont {C.}~\bibnamefont {Bustamante}},\ }\bibfield
   {title} {\bibinfo {title} {Overstretching {B-DNA}: The elastic response of
  individual double-stranded and single-stranded {DNA} molecules},\ }\href
  {https://doi.org/10.1126/science.271.5250.795} {\bibfield  {journal}
  {\bibinfo  {journal} {Science}\ }\textbf {\bibinfo {volume} {271}},\ \bibinfo
  {pages} {795} (\bibinfo {year} {1996})}\BibitemShut {NoStop}%
\bibitem [{\citenamefont {Oesterhelt}\ \emph {et~al.}(1999)\citenamefont
  {Oesterhelt}, \citenamefont {Rief},\ and\ \citenamefont
  {Gaub}}]{oesterhelt1999}%
  \BibitemOpen
  \bibfield  {author} {\bibinfo {author} {\bibfnamefont {F.}~\bibnamefont
  {Oesterhelt}}, \bibinfo {author} {\bibfnamefont {M.}~\bibnamefont {Rief}},\
  and\ \bibinfo {author} {\bibfnamefont {H.~E.}\ \bibnamefont {Gaub}},\
  }\bibfield  {title} {\bibinfo {title} {Single molecule force spectroscopy by
  {AFM} indicates helical structure of poly(ethylene-glycol) in water},\ }\href
  {https://doi.org/10.1088/1367-2630/1/1/006} {\bibfield  {journal} {\bibinfo
  {journal} {New Journal of Physics}\ }\textbf {\bibinfo {volume} {1}},\
  \bibinfo {pages} {6} (\bibinfo {year} {1999})}\BibitemShut {NoStop}%
\bibitem [{\citenamefont {Grebikova}\ \emph {et~al.}(2016)\citenamefont
  {Grebikova}, \citenamefont {Radiom}, \citenamefont {Maroni}, \citenamefont
  {Schlüter},\ and\ \citenamefont {Borkovec}}]{grebikova2016350}%
  \BibitemOpen
  \bibfield  {author} {\bibinfo {author} {\bibfnamefont {L.}~\bibnamefont
  {Grebikova}}, \bibinfo {author} {\bibfnamefont {M.}~\bibnamefont {Radiom}},
  \bibinfo {author} {\bibfnamefont {P.}~\bibnamefont {Maroni}}, \bibinfo
  {author} {\bibfnamefont {A.~D.}\ \bibnamefont {Schlüter}},\ and\ \bibinfo
  {author} {\bibfnamefont {M.}~\bibnamefont {Borkovec}},\ }\bibfield  {title}
  {\bibinfo {title} {Recording stretching response of single polymer chains
  adsorbed on solid substrates},\ }\href
  {https://doi.org/10.1016/j.polymer.2016.02.045} {\bibfield  {journal}
  {\bibinfo  {journal} {Polymer}\ }\textbf {\bibinfo {volume} {102}},\ \bibinfo
  {pages} {350} (\bibinfo {year} {2016})}\BibitemShut {NoStop}%
\bibitem [{\citenamefont {Grebikova}\ \emph {et~al.}(2014)\citenamefont
  {Grebikova}, \citenamefont {Maroni}, \citenamefont {Zhang}, \citenamefont
  {Schl{\"u}ter},\ and\ \citenamefont {Borkovec}}]{grebikova2014single}%
  \BibitemOpen
  \bibfield  {author} {\bibinfo {author} {\bibfnamefont {L.}~\bibnamefont
  {Grebikova}}, \bibinfo {author} {\bibfnamefont {P.}~\bibnamefont {Maroni}},
  \bibinfo {author} {\bibfnamefont {B.}~\bibnamefont {Zhang}}, \bibinfo
  {author} {\bibfnamefont {A.~D.}\ \bibnamefont {Schl{\"u}ter}},\ and\ \bibinfo
  {author} {\bibfnamefont {M.}~\bibnamefont {Borkovec}},\ }\bibfield  {title}
  {\bibinfo {title} {Single-molecule force measurements by nano-handling of
  individual dendronized polymers},\ }\href {https://doi.org/10.1021/nn405485h}
  {\bibfield  {journal} {\bibinfo  {journal} {ACS nano}\ }\textbf {\bibinfo
  {volume} {8}},\ \bibinfo {pages} {2237} (\bibinfo {year} {2014})}\BibitemShut
  {NoStop}%
\bibitem [{\citenamefont {Rief}\ \emph {et~al.}(1997)\citenamefont {Rief},
  \citenamefont {Oesterhelt}, \citenamefont {Heymann},\ and\ \citenamefont
  {Gaub}}]{rief1997single}%
  \BibitemOpen
  \bibfield  {author} {\bibinfo {author} {\bibfnamefont {M.}~\bibnamefont
  {Rief}}, \bibinfo {author} {\bibfnamefont {F.}~\bibnamefont {Oesterhelt}},
  \bibinfo {author} {\bibfnamefont {B.}~\bibnamefont {Heymann}},\ and\ \bibinfo
  {author} {\bibfnamefont {H.~E.}\ \bibnamefont {Gaub}},\ }\bibfield  {title}
  {\bibinfo {title} {Single molecule force spectroscopy on polysaccharides by
  atomic force microscopy},\ }\href
  {https://doi.org/10.1126/science.275.5304.1295} {\bibfield  {journal}
  {\bibinfo  {journal} {Science}\ }\textbf {\bibinfo {volume} {275}},\ \bibinfo
  {pages} {1295} (\bibinfo {year} {1997})}\BibitemShut {NoStop}%
\bibitem [{\citenamefont {Janshoff}\ \emph {et~al.}(2000)\citenamefont
  {Janshoff}, \citenamefont {Neitzert}, \citenamefont {Oberd{\"o}rfer},\ and\
  \citenamefont {Fuchs}}]{janshoff2000force}%
  \BibitemOpen
  \bibfield  {author} {\bibinfo {author} {\bibfnamefont {A.}~\bibnamefont
  {Janshoff}}, \bibinfo {author} {\bibfnamefont {M.}~\bibnamefont {Neitzert}},
  \bibinfo {author} {\bibfnamefont {Y.}~\bibnamefont {Oberd{\"o}rfer}},\ and\
  \bibinfo {author} {\bibfnamefont {H.}~\bibnamefont {Fuchs}},\ }\bibfield
  {title} {\bibinfo {title} {Force spectroscopy of molecular systems—single
  molecule spectroscopy of polymers and biomolecules},\ }\href
  {https://doi.org/10.1002/1521-3773(20000915)39:18<3212::aid-anie3212>3.0.co;2-x}
  {\bibfield  {journal} {\bibinfo  {journal} {Angewandte Chemie International
  Edition}\ }\textbf {\bibinfo {volume} {39}},\ \bibinfo {pages} {3212}
  (\bibinfo {year} {2000})}\BibitemShut {NoStop}%
\bibitem [{\citenamefont {Frey}\ \emph {et~al.}(2012)\citenamefont {Frey},
  \citenamefont {Gooding}, \citenamefont {Wijeratne},\ and\ \citenamefont
  {Kiang}}]{frey2012understanding}%
  \BibitemOpen
  \bibfield  {author} {\bibinfo {author} {\bibfnamefont {E.~W.}\ \bibnamefont
  {Frey}}, \bibinfo {author} {\bibfnamefont {A.~A.}\ \bibnamefont {Gooding}},
  \bibinfo {author} {\bibfnamefont {S.}~\bibnamefont {Wijeratne}},\ and\
  \bibinfo {author} {\bibfnamefont {C.-H.}\ \bibnamefont {Kiang}},\ }\bibfield
  {title} {\bibinfo {title} {Understanding the physics of {DNA} using nanoscale
  single-molecule manipulation},\ }\href
  {https://doi.org/10.1007/s11467-012-0261-0} {\bibfield  {journal} {\bibinfo
  {journal} {Frontiers of physics}\ }\textbf {\bibinfo {volume} {7}},\ \bibinfo
  {pages} {576} (\bibinfo {year} {2012})}\BibitemShut {NoStop}%
\bibitem [{\citenamefont {Wang}\ \emph {et~al.}(1997)\citenamefont {Wang},
  \citenamefont {Yin}, \citenamefont {Landick}, \citenamefont {Gelles},\ and\
  \citenamefont {Block}}]{WANG19971335}%
  \BibitemOpen
  \bibfield  {author} {\bibinfo {author} {\bibfnamefont {M.}~\bibnamefont
  {Wang}}, \bibinfo {author} {\bibfnamefont {H.}~\bibnamefont {Yin}}, \bibinfo
  {author} {\bibfnamefont {R.}~\bibnamefont {Landick}}, \bibinfo {author}
  {\bibfnamefont {J.}~\bibnamefont {Gelles}},\ and\ \bibinfo {author}
  {\bibfnamefont {S.}~\bibnamefont {Block}},\ }\bibfield  {title} {\bibinfo
  {title} {Stretching {DNA} with optical tweezers},\ }\href
  {https://doi.org/10.1016/S0006-3495(97)78780-0} {\bibfield  {journal}
  {\bibinfo  {journal} {Biophysical Journal}\ }\textbf {\bibinfo {volume}
  {72}},\ \bibinfo {pages} {1335} (\bibinfo {year} {1997})}\BibitemShut
  {NoStop}%
\bibitem [{\citenamefont {Wang}\ \emph {et~al.}(2001)\citenamefont {Wang},
  \citenamefont {Forbes},\ and\ \citenamefont {Jin}}]{wang2001single}%
  \BibitemOpen
  \bibfield  {author} {\bibinfo {author} {\bibfnamefont {K.}~\bibnamefont
  {Wang}}, \bibinfo {author} {\bibfnamefont {J.~G.}\ \bibnamefont {Forbes}},\
  and\ \bibinfo {author} {\bibfnamefont {A.~J.}\ \bibnamefont {Jin}},\
  }\bibfield  {title} {\bibinfo {title} {Single molecule measurements of titin
  elasticity},\ }\href {https://doi.org/10.1016/S0079-6107(01)00009-8}
  {\bibfield  {journal} {\bibinfo  {journal} {Progress in biophysics and
  molecular biology}\ }\textbf {\bibinfo {volume} {77}},\ \bibinfo {pages} {1}
  (\bibinfo {year} {2001})}\BibitemShut {NoStop}%
\bibitem [{\citenamefont {Calderon}\ \emph {et~al.}(2008)\citenamefont
  {Calderon}, \citenamefont {Chen}, \citenamefont {Lin}, \citenamefont
  {Harris},\ and\ \citenamefont {Kiang}}]{calderon2008quantifying}%
  \BibitemOpen
  \bibfield  {author} {\bibinfo {author} {\bibfnamefont {C.~P.}\ \bibnamefont
  {Calderon}}, \bibinfo {author} {\bibfnamefont {W.-H.}\ \bibnamefont {Chen}},
  \bibinfo {author} {\bibfnamefont {K.-J.}\ \bibnamefont {Lin}}, \bibinfo
  {author} {\bibfnamefont {N.~C.}\ \bibnamefont {Harris}},\ and\ \bibinfo
  {author} {\bibfnamefont {C.-H.}\ \bibnamefont {Kiang}},\ }\bibfield  {title}
  {\bibinfo {title} {Quantifying {DNA} melting transitions using
  single-molecule force spectroscopy},\ }\href
  {https://doi.org/10.1088/0953-8984/21/3/034114} {\bibfield  {journal}
  {\bibinfo  {journal} {Journal of Physics: Condensed Matter}\ }\textbf
  {\bibinfo {volume} {21}},\ \bibinfo {pages} {034114} (\bibinfo {year}
  {2008})}\BibitemShut {NoStop}%
\bibitem [{\citenamefont {McCauley}\ and\ \citenamefont
  {Williams}(2007)}]{McCauley2007}%
  \BibitemOpen
  \bibfield  {author} {\bibinfo {author} {\bibfnamefont {M.~J.}\ \bibnamefont
  {McCauley}}\ and\ \bibinfo {author} {\bibfnamefont {M.~C.}\ \bibnamefont
  {Williams}},\ }\bibfield  {title} {\bibinfo {title} {Mechanisms of {DNA}
  binding determined in optical tweezers experiments},\ }\href
  {https://doi.org/10.1002/bip.20622} {\bibfield  {journal} {\bibinfo
  {journal} {Biopolymers}\ }\textbf {\bibinfo {volume} {85}},\ \bibinfo {pages}
  {154} (\bibinfo {year} {2007})}\BibitemShut {NoStop}%
\bibitem [{\citenamefont {Bosco}\ \emph {et~al.}(2013)\citenamefont {Bosco},
  \citenamefont {Camunas-Soler},\ and\ \citenamefont {Ritort}}]{Bosco2013}%
  \BibitemOpen
  \bibfield  {author} {\bibinfo {author} {\bibfnamefont {A.}~\bibnamefont
  {Bosco}}, \bibinfo {author} {\bibfnamefont {J.}~\bibnamefont
  {Camunas-Soler}},\ and\ \bibinfo {author} {\bibfnamefont {F.}~\bibnamefont
  {Ritort}},\ }\bibfield  {title} {\bibinfo {title} {Elastic properties and
  secondary structure formation of single-stranded {DNA} at monovalent and
  divalent salt conditions},\ }\href {https://doi.org/10.1093/nar/gkt1089}
  {\bibfield  {journal} {\bibinfo  {journal} {Nucleic Acids Research}\ }\textbf
  {\bibinfo {volume} {42}},\ \bibinfo {pages} {2064} (\bibinfo {year}
  {2013})}\BibitemShut {NoStop}%
\bibitem [{\citenamefont {Camunas-Soler}\ \emph {et~al.}(2016)\citenamefont
  {Camunas-Soler}, \citenamefont {Ribezzi-Crivellari},\ and\ \citenamefont
  {Ritort}}]{camunas2016}%
  \BibitemOpen
  \bibfield  {author} {\bibinfo {author} {\bibfnamefont {J.}~\bibnamefont
  {Camunas-Soler}}, \bibinfo {author} {\bibfnamefont {M.}~\bibnamefont
  {Ribezzi-Crivellari}},\ and\ \bibinfo {author} {\bibfnamefont
  {F.}~\bibnamefont {Ritort}},\ }\bibfield  {title} {\bibinfo {title} {Elastic
  properties of nucleic acids by single-molecule force spectroscopy},\ }\href
  {https://doi.org/10.1146/annurev-biophys-062215-011158} {\bibfield  {journal}
  {\bibinfo  {journal} {Annual Review of Biophysics}\ }\textbf {\bibinfo
  {volume} {45}},\ \bibinfo {pages} {65} (\bibinfo {year} {2016})}\BibitemShut
  {NoStop}%
\bibitem [{\citenamefont {Fiasconaro}\ and\ \citenamefont
  {Falo}(2019)}]{fiasconaro2019analytical}%
  \BibitemOpen
  \bibfield  {author} {\bibinfo {author} {\bibfnamefont {A.}~\bibnamefont
  {Fiasconaro}}\ and\ \bibinfo {author} {\bibfnamefont {F.}~\bibnamefont
  {Falo}},\ }\bibfield  {title} {\bibinfo {title} {Analytical results of the
  extensible freely jointed chain model},\ }\href
  {https://doi.org/10.1016/j.physa.2019.121929} {\bibfield  {journal} {\bibinfo
   {journal} {Physica A: Statistical Mechanics and its Applications}\ }\textbf
  {\bibinfo {volume} {532}},\ \bibinfo {pages} {121929} (\bibinfo {year}
  {2019})}\BibitemShut {NoStop}%
\bibitem [{\citenamefont {Radiom}\ and\ \citenamefont
  {Borkovec}(2017)}]{radiom2017influence}%
  \BibitemOpen
  \bibfield  {author} {\bibinfo {author} {\bibfnamefont {M.}~\bibnamefont
  {Radiom}}\ and\ \bibinfo {author} {\bibfnamefont {M.}~\bibnamefont
  {Borkovec}},\ }\bibfield  {title} {\bibinfo {title} {Influence of
  ligand-receptor interactions on force-extension behavior within the freely
  jointed chain model},\ }\href {https://doi.org/10.1103/PhysRevE.96.062501}
  {\bibfield  {journal} {\bibinfo  {journal} {Phys. Rev. E}\ }\textbf {\bibinfo
  {volume} {96}},\ \bibinfo {pages} {062501} (\bibinfo {year}
  {2017})}\BibitemShut {NoStop}%
\bibitem [{\citenamefont {Jayathilaka}\ \emph {et~al.}(2021)\citenamefont
  {Jayathilaka}, \citenamefont {Molley}, \citenamefont {Huang}, \citenamefont
  {Islam}, \citenamefont {Buche}, \citenamefont {Silberstein}, \citenamefont
  {Kruzic},\ and\ \citenamefont {Kilian}}]{jayathilaka2021force}%
  \BibitemOpen
  \bibfield  {author} {\bibinfo {author} {\bibfnamefont {P.~B.}\ \bibnamefont
  {Jayathilaka}}, \bibinfo {author} {\bibfnamefont {T.~G.}\ \bibnamefont
  {Molley}}, \bibinfo {author} {\bibfnamefont {Y.}~\bibnamefont {Huang}},
  \bibinfo {author} {\bibfnamefont {M.~S.}\ \bibnamefont {Islam}}, \bibinfo
  {author} {\bibfnamefont {M.~R.}\ \bibnamefont {Buche}}, \bibinfo {author}
  {\bibfnamefont {M.~N.}\ \bibnamefont {Silberstein}}, \bibinfo {author}
  {\bibfnamefont {J.~J.}\ \bibnamefont {Kruzic}},\ and\ \bibinfo {author}
  {\bibfnamefont {K.~A.}\ \bibnamefont {Kilian}},\ }\bibfield  {title}
  {\bibinfo {title} {Force-mediated molecule release from double network
  hydrogels},\ }\href {https://doi.org/10.1039/D1CC02726C} {\bibfield
  {journal} {\bibinfo  {journal} {Chemical Communications}\ }\textbf {\bibinfo
  {volume} {57}},\ \bibinfo {pages} {8484} (\bibinfo {year}
  {2021})}\BibitemShut {NoStop}%
\bibitem [{\citenamefont {Mao}\ \emph {et~al.}(2017)\citenamefont {Mao},
  \citenamefont {Talamini},\ and\ \citenamefont {Anand}}]{mao2017rupture}%
  \BibitemOpen
  \bibfield  {author} {\bibinfo {author} {\bibfnamefont {Y.}~\bibnamefont
  {Mao}}, \bibinfo {author} {\bibfnamefont {B.}~\bibnamefont {Talamini}},\ and\
  \bibinfo {author} {\bibfnamefont {L.}~\bibnamefont {Anand}},\ }\bibfield
  {title} {\bibinfo {title} {Rupture of polymers by chain scission},\ }\href
  {https://doi.org/10.1016/j.eml.2017.01.003} {\bibfield  {journal} {\bibinfo
  {journal} {Extreme Mechanics Letters}\ }\textbf {\bibinfo {volume} {13}},\
  \bibinfo {pages} {17} (\bibinfo {year} {2017})}\BibitemShut {NoStop}%
\bibitem [{\citenamefont {Talamini}\ \emph {et~al.}(2018)\citenamefont
  {Talamini}, \citenamefont {Mao},\ and\ \citenamefont
  {Anand}}]{talamini2018progressive}%
  \BibitemOpen
  \bibfield  {author} {\bibinfo {author} {\bibfnamefont {B.}~\bibnamefont
  {Talamini}}, \bibinfo {author} {\bibfnamefont {Y.}~\bibnamefont {Mao}},\ and\
  \bibinfo {author} {\bibfnamefont {L.}~\bibnamefont {Anand}},\ }\bibfield
  {title} {\bibinfo {title} {Progressive damage and rupture in polymers},\
  }\href {https://doi.org/10.1016/j.jmps.2017.11.013} {\bibfield  {journal}
  {\bibinfo  {journal} {Journal of the Mechanics and Physics of Solids}\
  }\textbf {\bibinfo {volume} {111}},\ \bibinfo {pages} {434} (\bibinfo {year}
  {2018})}\BibitemShut {NoStop}%
\bibitem [{\citenamefont {Mao}\ and\ \citenamefont
  {Anand}(2018)}]{mao2018theory}%
  \BibitemOpen
  \bibfield  {author} {\bibinfo {author} {\bibfnamefont {Y.}~\bibnamefont
  {Mao}}\ and\ \bibinfo {author} {\bibfnamefont {L.}~\bibnamefont {Anand}},\
  }\bibfield  {title} {\bibinfo {title} {A theory for fracture of polymeric
  gels},\ }\href {https://doi.org/10.1016/j.jmps.2018.02.008} {\bibfield
  {journal} {\bibinfo  {journal} {Journal of the Mechanics and Physics of
  Solids}\ }\textbf {\bibinfo {volume} {115}},\ \bibinfo {pages} {30} (\bibinfo
  {year} {2018})}\BibitemShut {NoStop}%
\bibitem [{\citenamefont {Li}\ and\ \citenamefont
  {Bouklas}(2020)}]{li2020variational}%
  \BibitemOpen
  \bibfield  {author} {\bibinfo {author} {\bibfnamefont {B.}~\bibnamefont
  {Li}}\ and\ \bibinfo {author} {\bibfnamefont {N.}~\bibnamefont {Bouklas}},\
  }\bibfield  {title} {\bibinfo {title} {A variational phase-field model for
  brittle fracture in polydisperse elastomer networks},\ }\href
  {https://doi.org/10.1016/j.ijsolstr.2019.08.012} {\bibfield  {journal}
  {\bibinfo  {journal} {International Journal of Solids and Structures}\
  }\textbf {\bibinfo {volume} {182}},\ \bibinfo {pages} {193} (\bibinfo {year}
  {2020})}\BibitemShut {NoStop}%
\bibitem [{\citenamefont {Mulderrig}\ \emph {et~al.}(2021)\citenamefont
  {Mulderrig}, \citenamefont {Li},\ and\ \citenamefont
  {Bouklas}}]{mulderrig2021affine}%
  \BibitemOpen
  \bibfield  {author} {\bibinfo {author} {\bibfnamefont {J.}~\bibnamefont
  {Mulderrig}}, \bibinfo {author} {\bibfnamefont {B.}~\bibnamefont {Li}},\ and\
  \bibinfo {author} {\bibfnamefont {N.}~\bibnamefont {Bouklas}},\ }\bibfield
  {title} {\bibinfo {title} {Affine and non-affine microsphere models for chain
  scission in polydisperse elastomer networks},\ }\href
  {https://doi.org/10.1016/j.mechmat.2021.103857} {\bibfield  {journal}
  {\bibinfo  {journal} {Mechanics of Materials}\ }\textbf {\bibinfo {volume}
  {160}},\ \bibinfo {pages} {103857} (\bibinfo {year} {2021})}\BibitemShut
  {NoStop}%
\bibitem [{\citenamefont {Arunachala}\ \emph {et~al.}(2021)\citenamefont
  {Arunachala}, \citenamefont {Rastak},\ and\ \citenamefont
  {Linder}}]{arunachala2021energy}%
  \BibitemOpen
  \bibfield  {author} {\bibinfo {author} {\bibfnamefont {P.~K.}\ \bibnamefont
  {Arunachala}}, \bibinfo {author} {\bibfnamefont {R.}~\bibnamefont {Rastak}},\
  and\ \bibinfo {author} {\bibfnamefont {C.}~\bibnamefont {Linder}},\
  }\bibfield  {title} {\bibinfo {title} {Energy based fracture initiation
  criterion for strain-crystallizing rubber-like materials with pre-existing
  cracks},\ }\href {https://doi.org/10.1016/j.jmps.2021.104617} {\bibfield
  {journal} {\bibinfo  {journal} {Journal of the Mechanics and Physics of
  Solids}\ }\textbf {\bibinfo {volume} {157}},\ \bibinfo {pages} {104617}
  (\bibinfo {year} {2021})}\BibitemShut {NoStop}%
\bibitem [{\citenamefont {Lamont}\ \emph {et~al.}(2021)\citenamefont {Lamont},
  \citenamefont {Mulderrig}, \citenamefont {Bouklas},\ and\ \citenamefont
  {Vernerey}}]{lamont2021rate}%
  \BibitemOpen
  \bibfield  {author} {\bibinfo {author} {\bibfnamefont {S.~C.}\ \bibnamefont
  {Lamont}}, \bibinfo {author} {\bibfnamefont {J.}~\bibnamefont {Mulderrig}},
  \bibinfo {author} {\bibfnamefont {N.}~\bibnamefont {Bouklas}},\ and\ \bibinfo
  {author} {\bibfnamefont {F.~J.}\ \bibnamefont {Vernerey}},\ }\bibfield
  {title} {\bibinfo {title} {Rate-dependent damage mechanics of polymer
  networks with reversible bonds},\ }\href
  {https://doi.org/10.1021/acs.macromol.1c01943} {\bibfield  {journal}
  {\bibinfo  {journal} {Macromolecules}\ }\textbf {\bibinfo {volume} {54}},\
  \bibinfo {pages} {10801} (\bibinfo {year} {2021})}\BibitemShut {NoStop}%
\bibitem [{\citenamefont {Arora}\ \emph {et~al.}(2020)\citenamefont {Arora},
  \citenamefont {Lin}, \citenamefont {Beech}, \citenamefont {Mochigase},
  \citenamefont {Wang},\ and\ \citenamefont {Olsen}}]{arora2020fracture}%
  \BibitemOpen
  \bibfield  {author} {\bibinfo {author} {\bibfnamefont {A.}~\bibnamefont
  {Arora}}, \bibinfo {author} {\bibfnamefont {T.-S.}\ \bibnamefont {Lin}},
  \bibinfo {author} {\bibfnamefont {H.~K.}\ \bibnamefont {Beech}}, \bibinfo
  {author} {\bibfnamefont {H.}~\bibnamefont {Mochigase}}, \bibinfo {author}
  {\bibfnamefont {R.}~\bibnamefont {Wang}},\ and\ \bibinfo {author}
  {\bibfnamefont {B.~D.}\ \bibnamefont {Olsen}},\ }\bibfield  {title} {\bibinfo
  {title} {Fracture of polymer networks containing topological defects},\
  }\href {https://doi.org/10.1021/acs.macromol.0c01038} {\bibfield  {journal}
  {\bibinfo  {journal} {Macromolecules}\ }\textbf {\bibinfo {volume} {53}},\
  \bibinfo {pages} {7346} (\bibinfo {year} {2020})}\BibitemShut {NoStop}%
\bibitem [{\citenamefont {Yang}\ \emph {et~al.}(2020)\citenamefont {Yang},
  \citenamefont {Liechti},\ and\ \citenamefont {Huang}}]{yang2020multiscale}%
  \BibitemOpen
  \bibfield  {author} {\bibinfo {author} {\bibfnamefont {T.}~\bibnamefont
  {Yang}}, \bibinfo {author} {\bibfnamefont {K.~M.}\ \bibnamefont {Liechti}},\
  and\ \bibinfo {author} {\bibfnamefont {R.}~\bibnamefont {Huang}},\ }\bibfield
   {title} {\bibinfo {title} {A multiscale cohesive zone model for
  rate-dependent fracture of interfaces},\ }\href
  {https://doi.org/10.1016/j.jmps.2020.104142} {\bibfield  {journal} {\bibinfo
  {journal} {Journal of the Mechanics and Physics of Solids}\ }\textbf
  {\bibinfo {volume} {145}},\ \bibinfo {pages} {104142} (\bibinfo {year}
  {2020})}\BibitemShut {NoStop}%
\bibitem [{\citenamefont {Xiao}\ \emph {et~al.}(2021)\citenamefont {Xiao},
  \citenamefont {Han}, \citenamefont {Zhong},\ and\ \citenamefont
  {Qu}}]{xiao2021modeling}%
  \BibitemOpen
  \bibfield  {author} {\bibinfo {author} {\bibfnamefont {R.}~\bibnamefont
  {Xiao}}, \bibinfo {author} {\bibfnamefont {N.}~\bibnamefont {Han}}, \bibinfo
  {author} {\bibfnamefont {D.}~\bibnamefont {Zhong}},\ and\ \bibinfo {author}
  {\bibfnamefont {S.}~\bibnamefont {Qu}},\ }\bibfield  {title} {\bibinfo
  {title} {Modeling the mechanical behaviors of multiple network elastomers},\
  }\href {https://doi.org/10.1016/j.mechmat.2021.103992} {\bibfield  {journal}
  {\bibinfo  {journal} {Mechanics of Materials}\ }\textbf {\bibinfo {volume}
  {161}},\ \bibinfo {pages} {103992} (\bibinfo {year} {2021})}\BibitemShut
  {NoStop}%
\bibitem [{\citenamefont {Lavoie}\ \emph {et~al.}(2019)\citenamefont {Lavoie},
  \citenamefont {Long},\ and\ \citenamefont {Tang}}]{lavoie2019modeling}%
  \BibitemOpen
  \bibfield  {author} {\bibinfo {author} {\bibfnamefont {S.~R.}\ \bibnamefont
  {Lavoie}}, \bibinfo {author} {\bibfnamefont {R.}~\bibnamefont {Long}},\ and\
  \bibinfo {author} {\bibfnamefont {T.}~\bibnamefont {Tang}},\ }\bibfield
  {title} {\bibinfo {title} {Modeling the mechanics of polymer chains with
  deformable and active bonds},\ }\href
  {https://doi.org/10.1021/acs.jpcb.9b09068} {\bibfield  {journal} {\bibinfo
  {journal} {The Journal of Physical Chemistry B}\ }\textbf {\bibinfo {volume}
  {124}},\ \bibinfo {pages} {253} (\bibinfo {year} {2019})}\BibitemShut
  {NoStop}%
\bibitem [{\citenamefont {Guo}\ and\ \citenamefont
  {Za{\"i}ri}(2021)}]{guo2021micromechanics}%
  \BibitemOpen
  \bibfield  {author} {\bibinfo {author} {\bibfnamefont {Q.}~\bibnamefont
  {Guo}}\ and\ \bibinfo {author} {\bibfnamefont {F.}~\bibnamefont
  {Za{\"i}ri}},\ }\bibfield  {title} {\bibinfo {title} {A micromechanics-based
  model for deformation-induced damage and failure in elastomeric media},\
  }\href {https://doi.org/10.1016/j.ijplas.2021.102976} {\bibfield  {journal}
  {\bibinfo  {journal} {International Journal of Plasticity}\ }\textbf
  {\bibinfo {volume} {140}},\ \bibinfo {pages} {102976} (\bibinfo {year}
  {2021})}\BibitemShut {NoStop}%
\bibitem [{\citenamefont {Zhao}\ \emph {et~al.}(2021)\citenamefont {Zhao},
  \citenamefont {Lei}, \citenamefont {Chen}, \citenamefont {Zhang},
  \citenamefont {Wang},\ and\ \citenamefont {Lei}}]{zhao2021multiscale}%
  \BibitemOpen
  \bibfield  {author} {\bibinfo {author} {\bibfnamefont {Z.}~\bibnamefont
  {Zhao}}, \bibinfo {author} {\bibfnamefont {H.}~\bibnamefont {Lei}}, \bibinfo
  {author} {\bibfnamefont {H.-S.}\ \bibnamefont {Chen}}, \bibinfo {author}
  {\bibfnamefont {Q.}~\bibnamefont {Zhang}}, \bibinfo {author} {\bibfnamefont
  {P.}~\bibnamefont {Wang}},\ and\ \bibinfo {author} {\bibfnamefont
  {M.}~\bibnamefont {Lei}},\ }\bibfield  {title} {\bibinfo {title} {A
  multiscale tensile failure model for double network elastomer composites},\
  }\href {https://doi.org/10.1016/j.mechmat.2021.104074} {\bibfield  {journal}
  {\bibinfo  {journal} {Mechanics of Materials}\ }\textbf {\bibinfo {volume}
  {163}},\ \bibinfo {pages} {104074} (\bibinfo {year} {2021})}\BibitemShut
  {NoStop}%
\bibitem [{\citenamefont {Arora}\ \emph {et~al.}(2022)\citenamefont {Arora},
  \citenamefont {Lin},\ and\ \citenamefont {Olsen}}]{arora2021coarse}%
  \BibitemOpen
  \bibfield  {author} {\bibinfo {author} {\bibfnamefont {A.}~\bibnamefont
  {Arora}}, \bibinfo {author} {\bibfnamefont {T.-S.}\ \bibnamefont {Lin}},\
  and\ \bibinfo {author} {\bibfnamefont {B.~D.}\ \bibnamefont {Olsen}},\
  }\bibfield  {title} {\bibinfo {title} {Coarse-grained simulations for
  fracture of polymer networks: Stress versus topological inhomogeneities},\
  }\href {https://doi.org/10.1021/acs.macromol.1c01689} {\bibfield  {journal}
  {\bibinfo  {journal} {Macromolecules}\ }\textbf {\bibinfo {volume} {55}},\
  \bibinfo {pages} {4} (\bibinfo {year} {2022})}\BibitemShut {NoStop}%
\bibitem [{\citenamefont {Buche}(2021)}]{buche2021fundamental}%
  \BibitemOpen
  \bibfield  {author} {\bibinfo {author} {\bibfnamefont {M.~R.}\ \bibnamefont
  {Buche}},\ }\emph {\bibinfo {title} {Fundamental Theories for the Mechanics
  of Polymer Chains and Networks}},\ \href
  {https://www.proquest.com/openview/620bd18d1bf93950b88a41aa62ebdb3c} {Ph.D.
  thesis},\ \bibinfo  {school} {Cornell University} (\bibinfo {year}
  {2021})\BibitemShut {NoStop}%
\bibitem [{\citenamefont {Morse}(1929)}]{morse1929diatomic}%
  \BibitemOpen
  \bibfield  {author} {\bibinfo {author} {\bibfnamefont {P.~M.}\ \bibnamefont
  {Morse}},\ }\bibfield  {title} {\bibinfo {title} {{Diatomic molecules
  according to the wave mechanics. II. Vibrational levels}},\ }\href
  {https://doi.org/10.1103/PhysRev.34.57} {\bibfield  {journal} {\bibinfo
  {journal} {Physical Review}\ }\textbf {\bibinfo {volume} {34}},\ \bibinfo
  {pages} {57} (\bibinfo {year} {1929})}\BibitemShut {NoStop}%
\bibitem [{\citenamefont {Jones}(1924)}]{jones1924determinationii}%
  \BibitemOpen
  \bibfield  {author} {\bibinfo {author} {\bibfnamefont {J.~E.}\ \bibnamefont
  {Jones}},\ }\bibfield  {title} {\bibinfo {title} {On the determination of
  molecular fields. {II. F}rom the equation of state of a gas},\ }\href
  {https://doi.org/10.1098/rspa.1924.0082} {\bibfield  {journal} {\bibinfo
  {journal} {Proceedings of the Royal Society of London. Series A, Containing
  Papers of a Mathematical and Physical Character}\ }\textbf {\bibinfo {volume}
  {106}},\ \bibinfo {pages} {463} (\bibinfo {year} {1924})}\BibitemShut
  {NoStop}%
\bibitem [{\citenamefont {Buche}\ and\ \citenamefont
  {Grutzik}(2022)}]{buchegrutzikufjc2022}%
  \BibitemOpen
  \bibfield  {author} {\bibinfo {author} {\bibfnamefont {M.~R.}\ \bibnamefont
  {Buche}}\ and\ \bibinfo {author} {\bibfnamefont {S.~J.}\ \bibnamefont
  {Grutzik}},\ }\href {https://doi.org/10.5281/zenodo.6114263} {\bibinfo
  {title} {\texttt{ufjc}: {the Python package for the uFJC single-chain
  model}}},\ \bibinfo {howpublished} {Zenodo} (\bibinfo {year}
  {2022})\BibitemShut {NoStop}%
\bibitem [{\citenamefont {Bleistein}\ and\ \citenamefont
  {Handelsman}(1975)}]{bleistein1975asymptotic}%
  \BibitemOpen
  \bibfield  {author} {\bibinfo {author} {\bibfnamefont {N.}~\bibnamefont
  {Bleistein}}\ and\ \bibinfo {author} {\bibfnamefont {R.~A.}\ \bibnamefont
  {Handelsman}},\ }\href {https://store.doverpublications.com/0486650820.html}
  {\emph {\bibinfo {title} {Asymptotic Expansions of Integrals}}}\ (\bibinfo
  {publisher} {Ardent Media},\ \bibinfo {year} {1975})\BibitemShut {NoStop}%
\bibitem [{\citenamefont {Bender}\ and\ \citenamefont
  {Orszag}(2013)}]{bender2013advanced}%
  \BibitemOpen
  \bibfield  {author} {\bibinfo {author} {\bibfnamefont {C.~M.}\ \bibnamefont
  {Bender}}\ and\ \bibinfo {author} {\bibfnamefont {S.~A.}\ \bibnamefont
  {Orszag}},\ }\href {https://doi.org/10.1007/978-1-4757-3069-2} {\emph
  {\bibinfo {title} {Advanced Mathematical Methods for Scientists and Engineers
  I: Asymptotic Methods and Perturbation Theory}}}\ (\bibinfo  {publisher}
  {Springer Science \& Business Media},\ \bibinfo {year} {2013})\BibitemShut
  {NoStop}%
\bibitem [{\citenamefont {Rief}\ \emph {et~al.}(1998)\citenamefont {Rief},
  \citenamefont {Fernandez},\ and\ \citenamefont {Gaub}}]{rief1998elastically}%
  \BibitemOpen
  \bibfield  {author} {\bibinfo {author} {\bibfnamefont {M.}~\bibnamefont
  {Rief}}, \bibinfo {author} {\bibfnamefont {J.~M.}\ \bibnamefont
  {Fernandez}},\ and\ \bibinfo {author} {\bibfnamefont {H.~E.}\ \bibnamefont
  {Gaub}},\ }\bibfield  {title} {\bibinfo {title} {Elastically coupled
  two-level systems as a model for biopolymer extensibility},\ }\href
  {https://doi.org/10.1103/PhysRevLett.81.4764} {\bibfield  {journal} {\bibinfo
   {journal} {Physical Review Letters}\ }\textbf {\bibinfo {volume} {81}},\
  \bibinfo {pages} {4764} (\bibinfo {year} {1998})}\BibitemShut {NoStop}%
\bibitem [{\citenamefont {Manca}\ \emph {et~al.}(2013)\citenamefont {Manca},
  \citenamefont {Giordano}, \citenamefont {Palla}, \citenamefont {Cleri},\ and\
  \citenamefont {Colombo}}]{manca2013two}%
  \BibitemOpen
  \bibfield  {author} {\bibinfo {author} {\bibfnamefont {F.}~\bibnamefont
  {Manca}}, \bibinfo {author} {\bibfnamefont {S.}~\bibnamefont {Giordano}},
  \bibinfo {author} {\bibfnamefont {P.~L.}\ \bibnamefont {Palla}}, \bibinfo
  {author} {\bibfnamefont {F.}~\bibnamefont {Cleri}},\ and\ \bibinfo {author}
  {\bibfnamefont {L.}~\bibnamefont {Colombo}},\ }\bibfield  {title} {\bibinfo
  {title} {Two-state theory of single-molecule stretching experiments},\ }\href
  {https://doi.org/10.1103/PhysRevE.87.032705} {\bibfield  {journal} {\bibinfo
  {journal} {Physical Review E}\ }\textbf {\bibinfo {volume} {87}},\ \bibinfo
  {pages} {032705} (\bibinfo {year} {2013})}\BibitemShut {NoStop}%
\bibitem [{\citenamefont {Giordano}(2018)}]{giordano2018helmholtz}%
  \BibitemOpen
  \bibfield  {author} {\bibinfo {author} {\bibfnamefont {S.}~\bibnamefont
  {Giordano}},\ }\bibfield  {title} {\bibinfo {title} {Helmholtz and gibbs
  ensembles, thermodynamic limit and bistability in polymer lattice models},\
  }\href {https://doi.org/10.1007/s00161-017-0615-5} {\bibfield  {journal}
  {\bibinfo  {journal} {Continuum Mechanics and Thermodynamics}\ }\textbf
  {\bibinfo {volume} {30}},\ \bibinfo {pages} {459} (\bibinfo {year}
  {2018})}\BibitemShut {NoStop}%
\bibitem [{\citenamefont {Giordano}(2017)}]{giordano2017spin}%
  \BibitemOpen
  \bibfield  {author} {\bibinfo {author} {\bibfnamefont {S.}~\bibnamefont
  {Giordano}},\ }\bibfield  {title} {\bibinfo {title} {Spin variable approach
  for the statistical mechanics of folding and unfolding chains},\ }\href
  {https://doi.org/10.1039/C7SM00882A} {\bibfield  {journal} {\bibinfo
  {journal} {Soft matter}\ }\textbf {\bibinfo {volume} {13}},\ \bibinfo {pages}
  {6877} (\bibinfo {year} {2017})}\BibitemShut {NoStop}%
\bibitem [{\citenamefont {Benedito}\ and\ \citenamefont
  {Giordano}(2018{\natexlab{a}})}]{benedito2018thermodynamics}%
  \BibitemOpen
  \bibfield  {author} {\bibinfo {author} {\bibfnamefont {M.}~\bibnamefont
  {Benedito}}\ and\ \bibinfo {author} {\bibfnamefont {S.}~\bibnamefont
  {Giordano}},\ }\bibfield  {title} {\bibinfo {title} {Thermodynamics of small
  systems with conformational transitions: The case of two-state freely jointed
  chains with extensible units},\ }\href {https://doi.org/10.1063/1.5026386}
  {\bibfield  {journal} {\bibinfo  {journal} {The Journal of Chemical Physics}\
  }\textbf {\bibinfo {volume} {149}},\ \bibinfo {pages} {054901} (\bibinfo
  {year} {2018}{\natexlab{a}})}\BibitemShut {NoStop}%
\bibitem [{\citenamefont {Benedito}\ and\ \citenamefont
  {Giordano}(2018{\natexlab{b}})}]{benedito2018isotensional}%
  \BibitemOpen
  \bibfield  {author} {\bibinfo {author} {\bibfnamefont {M.}~\bibnamefont
  {Benedito}}\ and\ \bibinfo {author} {\bibfnamefont {S.}~\bibnamefont
  {Giordano}},\ }\bibfield  {title} {\bibinfo {title} {Isotensional and
  isometric force-extension response of chains with bistable units and ising
  interactions},\ }\href {https://doi.org/10.1103/PhysRevE.98.052146}
  {\bibfield  {journal} {\bibinfo  {journal} {Physical Review E}\ }\textbf
  {\bibinfo {volume} {98}},\ \bibinfo {pages} {052146} (\bibinfo {year}
  {2018}{\natexlab{b}})}\BibitemShut {NoStop}%
\bibitem [{\citenamefont {Powers}\ and\ \citenamefont
  {Sen}(2015)}]{powers2015mathematical}%
  \BibitemOpen
  \bibfield  {author} {\bibinfo {author} {\bibfnamefont {J.~M.}\ \bibnamefont
  {Powers}}\ and\ \bibinfo {author} {\bibfnamefont {M.}~\bibnamefont {Sen}},\
  }\href {https://doi.org/10.1017/CBO9781139583442} {\emph {\bibinfo {title}
  {Mathematical Methods in Engineering}}}\ (\bibinfo  {publisher} {Cambridge
  University Press},\ \bibinfo {year} {2015})\BibitemShut {NoStop}%
\bibitem [{\citenamefont {Manca}\ \emph {et~al.}(2014)\citenamefont {Manca},
  \citenamefont {Giordano}, \citenamefont {Palla},\ and\ \citenamefont
  {Cleri}}]{manca2014equivalence}%
  \BibitemOpen
  \bibfield  {author} {\bibinfo {author} {\bibfnamefont {F.}~\bibnamefont
  {Manca}}, \bibinfo {author} {\bibfnamefont {S.}~\bibnamefont {Giordano}},
  \bibinfo {author} {\bibfnamefont {P.~L.}\ \bibnamefont {Palla}},\ and\
  \bibinfo {author} {\bibfnamefont {F.}~\bibnamefont {Cleri}},\ }\bibfield
  {title} {\bibinfo {title} {On the equivalence of thermodynamics ensembles for
  flexible polymer chains},\ }\href
  {https://doi.org/10.1016/j.physa.2013.10.042} {\bibfield  {journal} {\bibinfo
   {journal} {Physica A: Statistical Mechanics and its Applications}\ }\textbf
  {\bibinfo {volume} {395}},\ \bibinfo {pages} {154} (\bibinfo {year}
  {2014})}\BibitemShut {NoStop}%
\bibitem [{\citenamefont {Treloar}(1946)}]{treloar1946statistical}%
  \BibitemOpen
  \bibfield  {author} {\bibinfo {author} {\bibfnamefont {L.~R.~G.}\
  \bibnamefont {Treloar}},\ }\bibfield  {title} {\bibinfo {title} {The
  statistical length of long-chain molecules},\ }\href
  {https://doi.org/10.5254/1.3543237} {\bibfield  {journal} {\bibinfo
  {journal} {Transactions of the Faraday Society}\ }\textbf {\bibinfo {volume}
  {42}},\ \bibinfo {pages} {77} (\bibinfo {year} {1946})}\BibitemShut {NoStop}%
\bibitem [{\citenamefont {Wang}\ and\ \citenamefont
  {Guth}(1952)}]{wang1952statistical}%
  \BibitemOpen
  \bibfield  {author} {\bibinfo {author} {\bibfnamefont {M.~C.}\ \bibnamefont
  {Wang}}\ and\ \bibinfo {author} {\bibfnamefont {E.}~\bibnamefont {Guth}},\
  }\bibfield  {title} {\bibinfo {title} {Statistical theory of networks of
  non-gaussian flexible chains},\ }\href {https://doi.org/10.1063/1.1700682}
  {\bibfield  {journal} {\bibinfo  {journal} {The Journal of Chemical Physics}\
  }\textbf {\bibinfo {volume} {20}},\ \bibinfo {pages} {1144} (\bibinfo {year}
  {1952})}\BibitemShut {NoStop}%
\bibitem [{\citenamefont {Harris}\ \emph {et~al.}(2020)\citenamefont {Harris},
  \citenamefont {Millman}, \citenamefont {van~der Walt}, \citenamefont
  {Gommers}, \citenamefont {Virtanen}, \citenamefont {Cournapeau},
  \citenamefont {Wieser}, \citenamefont {Taylor}, \citenamefont {Berg},
  \citenamefont {Smith}, \citenamefont {Kern}, \citenamefont {Picus},
  \citenamefont {Hoyer}, \citenamefont {van Kerkwijk}, \citenamefont {Brett},
  \citenamefont {Haldane}, \citenamefont {Fernández~del Río}, \citenamefont
  {Wiebe}, \citenamefont {Peterson}, \citenamefont {Gérard-Marchant},
  \citenamefont {Sheppard}, \citenamefont {Reddy}, \citenamefont {Weckesser},
  \citenamefont {Abbasi}, \citenamefont {Gohlke},\ and\ \citenamefont
  {Oliphant}}]{numpy}%
  \BibitemOpen
  \bibfield  {author} {\bibinfo {author} {\bibfnamefont {C.~R.}\ \bibnamefont
  {Harris}}, \bibinfo {author} {\bibfnamefont {K.~J.}\ \bibnamefont {Millman}},
  \bibinfo {author} {\bibfnamefont {S.~J.}\ \bibnamefont {van~der Walt}},
  \bibinfo {author} {\bibfnamefont {R.}~\bibnamefont {Gommers}}, \bibinfo
  {author} {\bibfnamefont {P.}~\bibnamefont {Virtanen}}, \bibinfo {author}
  {\bibfnamefont {D.}~\bibnamefont {Cournapeau}}, \bibinfo {author}
  {\bibfnamefont {E.}~\bibnamefont {Wieser}}, \bibinfo {author} {\bibfnamefont
  {J.}~\bibnamefont {Taylor}}, \bibinfo {author} {\bibfnamefont
  {S.}~\bibnamefont {Berg}}, \bibinfo {author} {\bibfnamefont {N.~J.}\
  \bibnamefont {Smith}}, \bibinfo {author} {\bibfnamefont {R.}~\bibnamefont
  {Kern}}, \bibinfo {author} {\bibfnamefont {M.}~\bibnamefont {Picus}},
  \bibinfo {author} {\bibfnamefont {S.}~\bibnamefont {Hoyer}}, \bibinfo
  {author} {\bibfnamefont {M.~H.}\ \bibnamefont {van Kerkwijk}}, \bibinfo
  {author} {\bibfnamefont {M.}~\bibnamefont {Brett}}, \bibinfo {author}
  {\bibfnamefont {A.}~\bibnamefont {Haldane}}, \bibinfo {author} {\bibfnamefont
  {J.}~\bibnamefont {Fernández~del Río}}, \bibinfo {author} {\bibfnamefont
  {M.}~\bibnamefont {Wiebe}}, \bibinfo {author} {\bibfnamefont
  {P.}~\bibnamefont {Peterson}}, \bibinfo {author} {\bibfnamefont
  {P.}~\bibnamefont {Gérard-Marchant}}, \bibinfo {author} {\bibfnamefont
  {K.}~\bibnamefont {Sheppard}}, \bibinfo {author} {\bibfnamefont
  {T.}~\bibnamefont {Reddy}}, \bibinfo {author} {\bibfnamefont
  {W.}~\bibnamefont {Weckesser}}, \bibinfo {author} {\bibfnamefont
  {H.}~\bibnamefont {Abbasi}}, \bibinfo {author} {\bibfnamefont
  {C.}~\bibnamefont {Gohlke}},\ and\ \bibinfo {author} {\bibfnamefont {T.~E.}\
  \bibnamefont {Oliphant}},\ }\bibfield  {title} {\bibinfo {title} {Array
  programming with \texttt{NumPy}},\ }\href
  {https://doi.org/10.1038/s41586-020-2649-2} {\bibfield  {journal} {\bibinfo
  {journal} {Nature}\ }\textbf {\bibinfo {volume} {585}},\ \bibinfo {pages}
  {357–362} (\bibinfo {year} {2020})}\BibitemShut {NoStop}%
\bibitem [{\citenamefont {Virtanen}\ \emph {et~al.}(2020)\citenamefont
  {Virtanen}, \citenamefont {Gommers}, \citenamefont {Oliphant}, \citenamefont
  {Haberland}, \citenamefont {Reddy}, \citenamefont {Cournapeau}, \citenamefont
  {Burovski}, \citenamefont {Peterson}, \citenamefont {Weckesser},
  \citenamefont {Bright}, \citenamefont {{van der Walt}}, \citenamefont
  {Brett}, \citenamefont {Wilson}, \citenamefont {Millman}, \citenamefont
  {Mayorov}, \citenamefont {Nelson}, \citenamefont {Jones}, \citenamefont
  {Kern}, \citenamefont {Larson}, \citenamefont {Carey}, \citenamefont {Polat},
  \citenamefont {Feng}, \citenamefont {Moore}, \citenamefont {{VanderPlas}},
  \citenamefont {Laxalde}, \citenamefont {Perktold}, \citenamefont {Cimrman},
  \citenamefont {Henriksen}, \citenamefont {Quintero}, \citenamefont {Harris},
  \citenamefont {Archibald}, \citenamefont {Ribeiro}, \citenamefont
  {Pedregosa}, \citenamefont {{van Mulbregt}},\ and\ \citenamefont {{SciPy 1.0
  Contributors}}}]{scipy}%
  \BibitemOpen
  \bibfield  {author} {\bibinfo {author} {\bibfnamefont {P.}~\bibnamefont
  {Virtanen}}, \bibinfo {author} {\bibfnamefont {R.}~\bibnamefont {Gommers}},
  \bibinfo {author} {\bibfnamefont {T.~E.}\ \bibnamefont {Oliphant}}, \bibinfo
  {author} {\bibfnamefont {M.}~\bibnamefont {Haberland}}, \bibinfo {author}
  {\bibfnamefont {T.}~\bibnamefont {Reddy}}, \bibinfo {author} {\bibfnamefont
  {D.}~\bibnamefont {Cournapeau}}, \bibinfo {author} {\bibfnamefont
  {E.}~\bibnamefont {Burovski}}, \bibinfo {author} {\bibfnamefont
  {P.}~\bibnamefont {Peterson}}, \bibinfo {author} {\bibfnamefont
  {W.}~\bibnamefont {Weckesser}}, \bibinfo {author} {\bibfnamefont
  {J.}~\bibnamefont {Bright}}, \bibinfo {author} {\bibfnamefont {S.~J.}\
  \bibnamefont {{van der Walt}}}, \bibinfo {author} {\bibfnamefont
  {M.}~\bibnamefont {Brett}}, \bibinfo {author} {\bibfnamefont
  {J.}~\bibnamefont {Wilson}}, \bibinfo {author} {\bibfnamefont {K.~J.}\
  \bibnamefont {Millman}}, \bibinfo {author} {\bibfnamefont {N.}~\bibnamefont
  {Mayorov}}, \bibinfo {author} {\bibfnamefont {A.~R.~J.}\ \bibnamefont
  {Nelson}}, \bibinfo {author} {\bibfnamefont {E.}~\bibnamefont {Jones}},
  \bibinfo {author} {\bibfnamefont {R.}~\bibnamefont {Kern}}, \bibinfo {author}
  {\bibfnamefont {E.}~\bibnamefont {Larson}}, \bibinfo {author} {\bibfnamefont
  {C.~J.}\ \bibnamefont {Carey}}, \bibinfo {author} {\bibfnamefont
  {{\.I}.}~\bibnamefont {Polat}}, \bibinfo {author} {\bibfnamefont
  {Y.}~\bibnamefont {Feng}}, \bibinfo {author} {\bibfnamefont {E.~W.}\
  \bibnamefont {Moore}}, \bibinfo {author} {\bibfnamefont {J.}~\bibnamefont
  {{VanderPlas}}}, \bibinfo {author} {\bibfnamefont {D.}~\bibnamefont
  {Laxalde}}, \bibinfo {author} {\bibfnamefont {J.}~\bibnamefont {Perktold}},
  \bibinfo {author} {\bibfnamefont {R.}~\bibnamefont {Cimrman}}, \bibinfo
  {author} {\bibfnamefont {I.}~\bibnamefont {Henriksen}}, \bibinfo {author}
  {\bibfnamefont {E.~A.}\ \bibnamefont {Quintero}}, \bibinfo {author}
  {\bibfnamefont {C.~R.}\ \bibnamefont {Harris}}, \bibinfo {author}
  {\bibfnamefont {A.~M.}\ \bibnamefont {Archibald}}, \bibinfo {author}
  {\bibfnamefont {A.~H.}\ \bibnamefont {Ribeiro}}, \bibinfo {author}
  {\bibfnamefont {F.}~\bibnamefont {Pedregosa}}, \bibinfo {author}
  {\bibfnamefont {P.}~\bibnamefont {{van Mulbregt}}},\ and\ \bibinfo {author}
  {\bibnamefont {{SciPy 1.0 Contributors}}},\ }\bibfield  {title} {\bibinfo
  {title} {\texttt{SciPy 1.0}: Fundamental algorithms for scientific computing
  in \texttt{Python}},\ }\href {https://doi.org/10.1038/s41592-019-0686-2}
  {\bibfield  {journal} {\bibinfo  {journal} {Nature Methods}\ }\textbf
  {\bibinfo {volume} {17}},\ \bibinfo {pages} {261} (\bibinfo {year}
  {2020})}\BibitemShut {NoStop}%
\bibitem [{\citenamefont {Inc.}()}]{Mathematica}%
  \BibitemOpen
  \bibfield  {author} {\bibinfo {author} {\bibfnamefont {W.~R.}\ \bibnamefont
  {Inc.}},\ }\href {https://www.wolfram.com/mathematica} {\bibinfo {title}
  {Mathematica, {V}ersion 12.3.1}},\ \bibinfo {note} {({Champaign, IL,
  2021)}}\BibitemShut {NoStop}%
\bibitem [{\citenamefont {Hinch}(1991)}]{hinch1991perturbation}%
  \BibitemOpen
  \bibfield  {author} {\bibinfo {author} {\bibfnamefont {E.~J.}\ \bibnamefont
  {Hinch}},\ }\href {https://doi.org/10.1017/CBO9781139172189} {\emph {\bibinfo
  {title} {Perturbation Methods}}}\ (\bibinfo  {publisher} {Cambridge
  University Press},\ \bibinfo {year} {1991})\BibitemShut {NoStop}%
\bibitem [{\citenamefont {Neumann}(2003)}]{neumann2003precise}%
  \BibitemOpen
  \bibfield  {author} {\bibinfo {author} {\bibfnamefont {R.~M.}\ \bibnamefont
  {Neumann}},\ }\bibfield  {title} {\bibinfo {title} {On the precise meaning of
  extension in the interpretation of polymer-chain stretching experiments},\
  }\href {https://doi.org/10.1016/S0006-3495(03)74760-2} {\bibfield  {journal}
  {\bibinfo  {journal} {Biophysical journal}\ }\textbf {\bibinfo {volume}
  {85}},\ \bibinfo {pages} {3418} (\bibinfo {year} {2003})}\BibitemShut
  {NoStop}%
\bibitem [{\citenamefont {S{\"u}zen}\ \emph {et~al.}(2009)\citenamefont
  {S{\"u}zen}, \citenamefont {Sega},\ and\ \citenamefont
  {Holm}}]{suzen2009ensemble}%
  \BibitemOpen
  \bibfield  {author} {\bibinfo {author} {\bibfnamefont {M.}~\bibnamefont
  {S{\"u}zen}}, \bibinfo {author} {\bibfnamefont {M.}~\bibnamefont {Sega}},\
  and\ \bibinfo {author} {\bibfnamefont {C.}~\bibnamefont {Holm}},\ }\bibfield
  {title} {\bibinfo {title} {Ensemble inequivalence in single-molecule
  experiments},\ }\href {https://doi.org/10.1103/PhysRevE.79.051118} {\bibfield
   {journal} {\bibinfo  {journal} {Physical Review E}\ }\textbf {\bibinfo
  {volume} {79}},\ \bibinfo {pages} {051118} (\bibinfo {year}
  {2009})}\BibitemShut {NoStop}%
\end{thebibliography}%

\end{document}